\documentclass[useAMS,usenatbib]{mn2e}

\usepackage{graphicx}

\title[Debris discs in NGC 1960]{Dust discs around intermediate mass
  and Sun-like stars in the 16 Myr old NGC 1960 open cluster}
\author[R. Smith and R. D. Jeffries]{R. Smith\thanks{E-mail:
    rs@astro.keele.ac.uk} and R.D. Jeffries \\
Astrophysics Group, Keele University, Keele, Staffordshire ST5 5BG}
\begin{document}

\date{In prep.}

\pagerange{\pageref{firstpage}--\pageref{lastpage}} \pubyear{2011}

\maketitle

\label{firstpage}

\begin{abstract}
We present an analysis of Spitzer IRAC (3.6--8$\mu$m) and MIPS
(24$\mu$m) imaging of members of the 16$^{+10}_{-5}$Myr old open
cluster NGC 1960 (M36).  Models of terrestrial planet formation
indicate that rocky planets are likely to achieve their final masses
at around 10--30Myr, and thus this cluster is at an interesting epoch
for planet formation.  We find 21 B-F5 type stars and 14 F6-K9 type
stars which have 24$\mu$m excess emission, and thus determine that
$>30$\% of B-F5 type stars and $>23$\% of F6-K9 type stars in this
cluster have 24$\mu$m excess emission.  These excess frequencies are
similar to those observed in other clusters of similar age.  Three
early type stars have excesses at near-infrared wavelengths.  Analysis
of their SEDs confirms that these are true debris discs and not
remnant primordial or transitional discs.  None of the 61 sun-like
stars have confirmed near-infrared excess, and we can place a limit on
the frequency of 8$\mu$m excess emission around sun-like stars of
$<$7\%. All of the detected excesses are consistent with emission from
debris discs and are not primordial.  
\end{abstract}

\begin{keywords}
circumstellar matter -- infrared: stars.
\end{keywords}

\section{Introduction}
  
Most young (1--2 Myr old) stars are surrounded by a gas-rich
primordial disc with dust producing high levels of emission in the infrared
($F_{IR}/F_\star \ge 0.1$, e.g. \citealt{kenyonhartmann}).  These are
the birthplaces of planets.  Current models of extra-solar planet formation
propose that dusty discs around a new star settle and km-sized
planetesimals aggregate on a short ($<$1Myr) timescale
\citep{weidenschillingcuzzi}.  The largest  planetestimals undergo
runaway accretion followed by oligarchic growth resulting in tens or
hundreds of 1000km-sized bodies in their own cleared ``feeding zones''
\citep{klahr08}.  These phases may take up to a few million 
years.  Circumstellar gas accretes onto the star or onto large
protoplanetary cores to create gas giant planets \citep{hartmann}.
Finally these planetary embryos collide and merge in a chaotic growth
phase to form a few stable terrestrial planets over 10-100Myr (see
e.g. \citealt{weidenschilling}).  These or other process remove the primordial
discs on timescales of 3--7Myr \citep{hernandez, hillenbrandreview,
  currie_ngc2362}.  By 10 Myr almost all remaining discs are optically
thin debris discs (see e.g. \citealt{wyattreview} and references
therein). Emission from such discs arises from second generation dust 
populations produced in collisions between planetesimals.  These dust
grains absorb and re-radiate star light at wavelengths typically
$>$10$\mu$m.   

The Spitzer Space telescope has been used to study the evolution of
debris discs in a statistical manner (see e.g. \citealt{rieke}, \citealt{su},
\citealt{siegler}, \citealt{rebull}, \citealt{carpenter}).  These
studies have tried to answer the question of why two apparently
similar stars can have very different levels of excess emission.  To
date, the clearest dependency is on age.  In A and early F-type stars
there is evidence for a peak in the upper envelope of 
excess emission at 10-20Myr before a decay in proportion with time
(see e.g. \citealt{wyattreview} and references therein).  For
solar-type stars the number of observed objects is smaller and so
correlations are harder to establish.  Based on current evidence the
decay of 24$\mu$m excess around solar-type stars appears to follow a
similar pattern to the A stars but on a timescale that is an order of
magnitude shorter (a drop from 40\% to 20\% of stars with 24$\mu$m
excess occurs between 10 and 100Myr for solar-type stars, and between
100 and 500Myr for A-type stars, see Figure 6 of \citealt{siegler}). 
In general the levels of excess emission are also smaller around lower
mass objects ($<$2 times the photospheric flux)
apart from around some of the youngest sources.  These results can be
interpreted within the framework of the evolutionary models of Kenyon
and Bromley (\citeyear{kenyon05,kenyon06}).  These suggest that 
planetesimals take longer to form at 10-30au around A-type stars than
at $\sim$1au around G-type stars (assuming thermal equilibrium, an excess
detected at $\ge$24$\mu$m implies a temperature of 100--150K, translating to
dust orbital radii of 3--30au around A and early F-type stars and 0.5--3au
around solar-type stars). Thus there are copious dust producing
collisional events for 10-100Myr during planetesimal accretion around
A-type stars.  Around solar-type stars 100km-sized embryos may be
complete within a few Myr, and thus subsequent observed debris is
likely to have been produced in recent catastrophic collisions like
the impact that formed the Earth-Moon system (see e.g. \citealt{canup}
and references therein).  These massive collisions are seen in as
peaks in the excess emission above a lower background of emission in 
simulations of planetesimal discs (see e.g. \citealt{kenyon05}).
Other authors have found these collisions could ignite collisional
avalanches in the disc, which are brighter in discs with high optical
depths (large populations of small grains, \citealt{grigorieva_avalanche}),
and could lead to longer-lived increased emission.  

The observed peak in debris disc emission for A and early F-type stars
(10--20Myr) is an important
epoch for terrestrial planet formation, as models and radiometric
dating suggest that terrestrial planets reach their final mass
within 10--30Myr \citep{wetherill, yin, kenyon06}.  Disc emission from this  
epoch could therefore provide a direct tracer of the frequency of
terrestrial planet-forming collisions.  Cluster samples at this age
have already been the target of Spitzer campaigns.  Clusters with
published 24$\mu$m Spitzer photometry include: the 16--17 Myr Sco-Cen
association ($\sim$40 stars across F and G spectral types, \citealt{chen}); 
 the 27Myr old IC 4665 cluster (73 stars, A to mid-K types
 \citealt{smith_ic4665}); NGC 2547 at 35--40 Myrs old
 (\citealt{naylor}; photometry for $\sim$70 sources); NGC 2232 at
 25Myr (209 sources B--M type but only 38 have published 24$\mu$m
 excess, \citealt{currie_ngc2232}); and h and $\chi$ Persei at 13 Myr 
(616 sources, spectral types B--G but uncertain membership lists,
 \citealt{currie_persei}).  It is difficult to obtain sufficient
 statistics from individual clusters to examine model predictions
 (either due to uncertain memberships or issues with obtaining
 sufficient 24$\mu$m photometry).  Combining open
cluster data is the primary method of studying factors such as the 
influence of stellar mass on the incidence of debris discs at the
epoch of terrestrial planet formation.  As open clusters provide a
homogeneous, chemically uniform coeval population, studies of multiple
open clusters allow tests of how environmental factors influence
debris emission, and the planet forming processes that create it.

NGC 1960 (M36) is a young open cluster, with an age determined by isochrone
fitting of 16$^{+10}_{-5}$Myr \citep{sanner}.  The large error bars on
this value are a reflection of the difficulty of determining the age of
this cluster as it has no known significantly evolved stars.  Other
recent age estimates in the literature include an age of $\sim$20Myr
\citep{mayne} and 25Myr \citep{hasan}.  As these are within the errors
given by \citet{sanner}, we adopt an age of 16$^{+10}_{-5}$Myr in this
paper. The cluster lies at a distance of 1318$\pm$120pc and has a
reddening of $E_{B-V} = 0.25\pm0.02$mag \citep{sanner}.  This cluster
provides an 
ideal complement to the studies of NGC 2232 and h and $\chi$ Persei as
it lies between these clusters in age.  Combining the studies of these
clusters will allow us to place constraints on the evolution of debris
discs within the peak of debris disc emission itself.  It is also
worth noting that this cluster can be viewed as an older analogue of
the Orion Nebula Cluster (ONC).  Integrating a Kroupa mass function
\citep{kroupa} and 
noting that there are $\sim$30 objects with masses of 3--7$M_\odot$
(using a distance modulus of $(m-M)_0=10.6\pm0.2$mag and extinction of
$A_V = 0.8$, \citealt{sanner}, and finding objects with absolute magnitudes
$\sim$-1.4--0.79, see target selection in next section) there are
$\sim$445 stars with masses 0.5--7$M_\odot$ and $\sim$1680 stars with
masses 0.1--7$M_\odot$.  These numbers are comparable with the ONC.
We would also expect 14 stars more massive than 7$M_\odot$ and find 16
stars of this size in NGC1960 (see target selection in the next
section), which is also comparable to the ONC.  

In this paper we present Spitzer archive data of NGC 1960.  We use
IRAC and MIPS 24$\mu$m photometry to identify excess emission around
cluster members based on $K_S - [24]$ vs $V - K_S$ colours and SED
fitting.  We discuss how levels of excess compare with other cluster
samples at this epoch, and how these observations constrain models of
terrestrial planet formation.

\section{NGC 1960 targets}

\begin{figure}
\includegraphics[width=7cm]{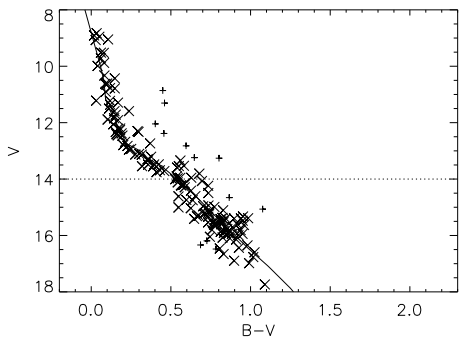}
\caption{\label{fig:bright_select} A $V$ vs $B-V$ colour-magnitude
  diagram of bright candidate members of NGC 1960 from the sample
  observed by \citet{sanner}.  Only targets with proper motions
  consistent with cluster membership are shown.  Proper motion is
  accurate for targets with $V<14$, which we adopt as a brightness
  cut-off in our target sample to avoid contamination by field
  sources.  A 16Myr-old isochrone from \citet{siess} is overplotted
  with conversion following \citet{kenyonhartmann} adjusted for a
  distance of 1300pc, $E(B-V)$=0.25.
  Targets marked by plus signs have colours inconsistent with cluster
  membership.  Crosses mark the sources with colours consistent with
  cluster membership.}
\end{figure}

To identify bright target members of the cluster we used the
sample presented by \citet{sanner} and adopt the same proper motion
and colour membership criteria presented in that paper.  From their initial
sample of 864 stars with $B$ and $V$ band photometry, 121 sources were
found to have proper motions consistent with cluster membership
(within 3$\sigma$ of the central cluster proper motion
$2.9\pm2.7$mas/yr in RA, $-8.0\pm2.5$mas/yr in Dec).  Proper motion
detections were considered accurate for sources with $V < 14$mag, and
thus we adopt this brightness as a cut-off for membership from this
catalogue ($V = 14$ translates to a spectral type of $\sim$G0 for this
cluster).
As a further test of cluster membership the $B-V$ versus $V$
colour-magnitude of the targets is compared to a 16Myr-old isochrone
from \citet{siess} with $E(B-V) =0.25$mag (as determined by isochrone
fitting in \citealt{sanner}).  The colour-magnitude diagram is shown
in Figure \ref{fig:bright_select}.  This led to the rejection of 7
targets with $V<14$ and a final selection of 63 targets.  We extract
$K_S$ magnitudes for the targets from the 2MASS catalogue
\citep{2MASS}.  This source list is given in Table \ref{tab:bright}.   

A set of lower mass members of NGC~1960 was assembled from spectroscopy
obtained at the William Herschel 4.2-m telescope. A separate paper is
being prepared using these data, but the details relevant to this paper
are given below. Observations were performed on 2009 November 25--26
and 2010 November 12--14 using the AF2/WYFFOS multifibre spectrograph,
with an echelle grating that provided a spectral coverage of 420\AA\ centred
at 6600\AA\ and a resolving power of 11\,000. Targets with $14<V<18.5$
were selected from a $V$ versus $V-I$ colour-magnitude diagram,
constructed using a $BVI$ photometric survey performed at the Isaac Newton
Telescope (N. Mayne private communication).  We took spectra of 472
unique objects (64 were observed twice or more) along with a suite of
standard stars.

The data were reduced and analysed as described in
\citet{jeffries_ic4665}. Bespoke software was used to flatfield,
extract and wavelength calibrate the spectra before cross-correlation
with radial velocity (RV) standards using the wavelength range
6370--6550\AA. We obtained good radial velocities for 358 unique
targets -- the precision varied with signal-to-noise ratio and
rotational broadening. In a cluster as young as NGC 1960, the presence
of measurable Li in the atmospheres of its cool stars is expected.
Indeed, Li is depleted so rapidly in low-mass stars that its presence
in K or M-type stars ($V-K_S > 2.0$) is a very strong indication of
youth ($<100$ Myr, \citealt{jeffries_book}).  The spectra were examined
for the presence of a 
lithium resonance line at 6708\AA. After correcting all the results to
a common heliocentric frame, it was clear that the Li-rich objects
defined a kinematic cluster within the data. After an iterative
clipping algorithm was applied we found that the Li-rich objects had a mean
heliocentric RV of $-4.50 \pm 0.14$\,km\,s$^{-1}$, with an additional
estimated zero point uncertainty of about 0.5\,km\,s$^{-1}$.  An
analysis of stars with two or more spectra showed that our RV
precision estimates are realistic.

To select members we chose to include all those Li-rich stars which
have a heliocentric RV within 2 error bars of the cluster mean, where
this error included an additional systematic 1\,km\,s$^{-1}$
representing a typical internal RV dispersion for a cluster (see
\citealt{jeffries_ic4665}). This gave a list of 69 candidate members with a
median RV precision of 1.3\,km\,s$^{-1}$ (targets are listed in Table
\ref{tab:sources}). To estimate the level of any
possible contamination, we make a similar selection from the Li-rich
stars, but using mean RVs
that are displaced by $\pm 15$\,km\,s$^{-1}$ from the actual cluster
mean. This resulted in the selection of $\sim 6$ objects which,
despite their discrepant RVs, could still be close-binary cluster
members.  We conclude that this sample of low-mass cluster candidates
has $<10$ per cent contamination.

For both the bright and lower mass samples temperatures are estimated
using the $V-K_S$ colours of the targets using the relation $$
5040/\rm{T}_{\rm{eff}} = 0.555 + 0.195(V-K_s) + 0.013(V-K_s)^2 $$
\citep{alonso}. Colours were first dereddened using $E(B-V) = 0.25$
\citep{sanner} and conversions to $E(V-K_S)$ from
\citet{rieke&lebofsky}.  For some of the higher mass targets (spectral
type earlier than F0) a relation in $B-V$ is more often favoured to
approximate temperature.  Using the relation $B-V =
-3.684\log \rm{T}_{\rm{eff}} +14.551$ we find temperatures that can be lower
by up to 10\% for the highest mass targets.  In this paper temperature
is only used to determine colour corrections on the photometry.
Within the $\sim 6000-11000$K temperature range of the higher mass
targets, this translates to a maximum difference in aperture
correction factor of $<$0.5\%.

\begin{table}
\caption{\label{tab:bright} Target sources in NGC 1960 from
  \citet{sanner}. Magnitudes and colours corrected for reddening
  (using $A_v =0.775$, $E(B-V) =0.25$; \citealt{sanner}, and
  $E(V-K_{\rm{S}})=0.686$ using the relations from
  \citealt{rieke&lebofsky}).  Only a few rows are shown here to
  illustrate the content, the full table is available online.}
\setlength{\tabcolsep}{4pt}
\begin{tabular}{lccrrrr} \hline ID & RA & Dec & $V_0$ &
  \small{($B$-$V$)$_0$} & \small{($V$-$K_{\rm{s}}$)$_0$} &
  \multicolumn{1}{c}{$T_{\rm{eff}}$, K} \\ \hline  
1 & 84.06585 & 34.1436 & 8.052 & -0.216 & -0.681 & 11769 \\ 
2 & 84.09608 & 34.1758 & 8.133 & -0.232 & -0.696 & 11842 \\ 
3 & 84.16359 & 34.0639 & 8.178 & -0.199 & -0.666 & 11696 \\ 
4 & 84.17627 & 34.2017 & 8.275 & -0.144 & -0.137 & 9535 \\ 
5 & 84.13335 & 34.1798 & 8.297 & -0.222 & -0.698 & 11852 \\ 
\ldots &     &         &       &        &        &      \\
\hline
\end{tabular}
\end{table}

\begin{table} \caption{\label{tab:sources} The lower mass target
    sources in NGC 1960.  Magnitudes and colours corrected for
    reddening (using $A_v =0.775$, $E(B-V) =0.25$; \citealt{sanner}, and
  $E(V-I)=0.400$ and $E(V-K_{\rm{S}})=0.686$ using the relations from
    \citealt{rieke&lebofsky}).  Only a few rows are shown here to
  illustrate the content, the full table is available online.} 
\setlength{\tabcolsep}{4pt}
\begin{tabular}{lccrrrr} \hline ID & RA & Dec & $V_0$ & ($V$-$I$)$_0$ &
  ($V$-$K_{\rm{s}}$)$_0$ & $T_{\rm{eff}}$, K \\ \hline 
1\_136 & 84.06671 & 34.0180 & 14.378 & 0.675 & 1.725 & 5419 \\ 
1\_138 & 84.04419 & 33.8800 & 14.050 & 0.426 & 1.033 & 6542 \\ 
1\_151 & 83.91997 & 33.9586 & 14.155 & 0.449 & 1.273 & 6114 \\ 
1\_223 & 84.09403 & 33.9216 & 14.999 & 0.664 & 1.788 & 5332 \\ 
1\_371 & 84.12115 & 34.0361 & 16.086 & 0.963 & 2.438 & 4550 \\ 
\ldots &     &         &       &        &        &      \\
\hline
\end{tabular}
\end{table}

\section{Spitzer data}

Data were sourced from the Spitzer Heritage Archive (SHA).  Data were
taken with the IRAC \citep{irac} and MIPS \citep{mips} instruments
under Spitzer Programme P50265.  IRAC data were obtained on 23rd
October 2008 in high dynamic range mode using a 6$\times$8 mapping
array.  The AOR consists of 260\arcsec map steps, array orientation and a
5-point Gaussian dither with small scale factor.   A frame time of 12
second exposures gave a 5$\sigma $ point-source sensitivity of 17.4,
16.6, 15.7 and 14.8 magnitudes in the four IRAC bands (3.6$\mu$m,
4.5$\mu$m, 5.8$\mu$m and 8$\mu$m) respectively.

MIPS data were obtained on 1st November 2008 in scan mapping mode
centered at 5h36m12s +34d08m24s.  A medium scan rate, scan leg length
of 0.5$^\circ$ and 11 scan legs were used to cover an area of
38\arcmin$\times$55\arcmin. Five repetitions of the AOR were obtained,
giving a 5$\sigma$ point-source sensitivity of 11.4 magnitudes at
24$\mu$m.   

The data were extracted as BCD (basic calibrated data) files from the
archive.  These data are individually flux-calibrated array images.
The Spitzer Science Center MOPEX package \citep{makovozmarleau} was used to
produce the final mosaics.  We used standard MOPEX modules.  The
individual 24$\mu$m MIPS frames were flat-fielded using the flatfield
module in MOPEX.  Overlap correction was determined using the default
settings in the overlap module and the final image mosaic consisting
of all five repetitions of the AOR was constructed using the mosaic
module.  Mosaics were created for each of the IRAC channels using the
overlap and mosaic modules in MOPEX with default settings.  For
details of these modules see \citet{makovozmarleau} or the on-line
MOPEX user's guide at
http://ssc.spitzer.caltech.edu/dataanalysistools/tools/
mopex/mopexusersguide/.  
Photometry was extracted using the APEX package from MOPEX.  For the
IRAC channels the PSF is undersampled and thus photometry was
extracted in circular apertures of radius 3 pixels ($\sim$3\farcs6)
with background determined in an annulus of inner radius 12 pixels and
outer radius 20 pixels ($\sim$14\farcs4--24\arcsec).  Apertures were
centered on the location of each source as listed in Tables
\ref{tab:bright} and \ref{tab:sources}.  Of the 132 targets in the
full sample, 4 (2, 4, 2) fell outside the channel 1 (2, 3, 4) imaging
mosaics.  Array location-dependent corrections were applied to the IRAC
photometry.  Correction images available on-line at
http://ssc.spitzer.caltech.edu/irac/calibrationfiles/locationcolor/
were mosaiced in the same way as the data frames to produce a
correction mosaic.  The aperture photometry of a source was then
multiplied by the value of the correction mosaic at the center of the
source image.  Aperture corrections were taken from tabulated values
listed in the IRAC data handbook.  Colour corrections were applied by
interpolation from tabulated values using the source effective
temperatures as listed in Tables \ref{tab:bright} and
\ref{tab:sources}.  We used the tabulated values in the IRAC data
handbook to convert the flux of the targets in Jy to magnitudes.
Specifically, the zero-points used were 280.9Jy at 3.6$\mu$m, 179.7Jy
at 4.5$\mu$m, 115.0Jy at 5.8$\mu$m and 64.1Jy at 8.0$\mu$m.  Absolute
calibration of IRAC is stable to 1--3\% \citep{reach2005}.  We add a
3\% calibration uncertainty in quadrature to statistical background
errors from pixel to pixel variation determined in the aperture module
to give a final error on the IRAC photometry.  The photometry was
de-reddened using the values determined by \citet{sanner} converted to
IRAC wavelengths using relations of \citet{flaherty}.  The final
photometry is listed in Tables \ref{tab:resultsbright} and
\ref{tab:results}.   

In the MIPS data the PSF is not undersampled.  We used the APEX
PRF(Point Response Function) fitting module to determine a PRF model
for the final mosiac using the brightest and cleanest stars (no
near neighbours, no bad pixels) in our target list to determine a PRF
model using the prf\_estimate module.
This model was used to fit the target stars to determine source
position (from input lists taken from Tables \ref{tab:bright} and
\ref{tab:sources}) and flux in the PRF photometry module in APEX.  If
the source could not be well fit (according to a $\chi^2$ analysis)
with either a single or multiple point sources (active deblend), then
the PRF photometry was determined to have failed.  For sources which
were not fit in the PRF photometry, or where the PRF fits were
sufficiently distant from the input source position that the detected
source was unlikely to be our target ($>$1\farcs225 corresponding to
1/5 of the FWHM of the average PSF for all sources) then the
aperture photometry was adopted instead. Apertures of radius 2.6
pixels ($\sim$6\farcs37) were used with annuli of inner radius 8.16
pixels and outer radius 13.06 pixels (20\arcsec -- 32\arcsec) to
determine the background.   

\begin{table*} 
\caption{\label{tab:resultsbright} Spitzer IRAC and MIPS
    24$\mu$m photometry for on bright members of NGC 1960.   Only a
    few rows are shown here to illustrate the content, the full table
    is available online.} 
\begin{tabular}{lrrrrrrrl} \hline Source & \multicolumn{1}{c}{[3.6]} &
    \multicolumn{1}{c}{[4.5]} & \multicolumn{1}{c}{[5.8]} &
    \multicolumn{1}{c}{[8.0]} & \multicolumn{1}{c}{[24]} &
    \multicolumn{1}{c}{$K_{\rm{s}}$ - [24]} &
    \multicolumn{1}{c}{$F_{\rm{24}}/F_{\rm{phot}}$} & Comments  \\ \hline 
1 & 8.909 [0.033] & 8.844 [0.033] & 8.876 [0.033] & 8.966 [0.033] &
8.918 [0.046] & -0.183 & 0.906 &  \\  
2 & 8.985 [0.033] & 8.953 [0.033] & 8.990 [0.033] & 9.061 [0.033] &
8.784 [0.045] & 0.048 & 1.122 &  \\  
3 & 8.968 [0.033] & 8.934 [0.033] & 8.993 [0.033] & 9.034 [0.033] &
9.129 [0.048] & -0.283 & 0.826 &  \\  
4 & 8.106 [0.033] & 7.709 [0.033] & 7.549 [0.033] & 7.310 [0.033] &
6.262 [0.043] & 2.152 & 7.424 & XS \\  
5 & 9.154 [0.033] & 9.110 [0.033] & 9.147 [0.033] & 9.221 [0.033] &
9.259 [0.048] & -0.261 & 0.845 &  \\  
\ldots &     &         &       &        &        &    & &  \\
33 & 11.140 [0.033] & 11.122 [0.033] & 11.186 [0.034] & 11.254 [0.038]
& $>$11.453 &  $<$-0.410 & $<$0.423 & Contam. \\  
\ldots &     &         &       &        &        &    & &  \\
\hline  \end{tabular}
\begin{flushleft}
\small{Notes: Stars which had contaminating flux of more than 13\%
  subtracted from the aperture photometry are marked as ``Contam'' in
  the comments column. None of these targets have been identified as a
  having confirm 24$\mu$m excess.  
  Sources with $K_S - [24]$ more than 3$\sigma$
  higher than the photospheric relation from \citet{plavchan} are
  marked as excess sources by the note ``XS'' in the comments column.
  The error on the photospheric relation includes differences from
  different photospheric predictions, variation in the colours
  observed for sources without excess and statistical error on the
  24$\mu$m photometry added in quadrature as described in the text. }
\end{flushleft}
\end{table*}

\begin{table*} \caption{\label{tab:results} Spitzer IRAC and MIPS
    24$\mu$m photometry for low-mass members of NGC 1960.   Only a
    few rows are shown here to illustrate the content, the full table
    is available online. }
\begin{tabular}{lccccrrrl} \hline Source & [3.6] & [4.5] & [5.8] & [8.0] &
  \multicolumn{1}{c}{[24]} & $K_{\rm{s}}$ - [24] &
  $F_{\rm{24}}/F_{\rm{phot}}$ & Comments  \\ \hline 
1\_136 & 12.621 [0.033] & 12.645 [0.033] & 12.679 [0.039] & 12.726
[0.055] & $>$11.940 &  $<$0.715 & $<$1.890 &  \\  
1\_138 &  -- & 12.935 [0.034] &  -- & 13.312 [0.119] & 11.886 [0.300]
& 1.134 & 2.851 & XS \\  
1\_151 & 12.809 [0.033] & 12.810 [0.034] & 12.832 [0.040] & 12.838
[0.084] & 11.385 [0.166] & 1.499 & 3.958 & XS \\  
1\_223 & 13.254 [0.034] & 13.272 [0.034] & 13.284 [0.050] & 13.508
[0.156] & $>$12.007 &  $<$1.206 & $<$2.964 &  \\  
1\_371 & 13.580 [0.034] & 13.486 [0.035] & 13.626 [0.058] & 13.828
[0.214] & $>$11.930 &  $<$1.720 & $<$4.623 &  \\  
\ldots &     &         &       &        &        &    & &  \\
\hline
\end{tabular}
\begin{flushleft}
\small{Notes: See notes on Table \ref{tab:resultsbright} for
  description of comments column. }
\end{flushleft}
\end{table*}

Some of our targets are close enough to other objects for
contamination of the aperture photometry (not PRF photometry) to be an
issue.  To determine a correction for contaminating sources in the
aperture photometry we used the APEX module in
MOPEX to create a list of all source detections ($>3\sigma$) in the
MIPS 24$\mu$m mosaic.  We identified the sources that were most
isolated from other detections (no other detection within 10
pixels, 25\arcsec of the source) and grouped them by brightness (PRF
determined flux).  We then used the aperture module to place apertures
(of the same size, 6\farcs37 radius, used for the target sources) at
the source and then at increasing distance from the source (in a
direction away from other detections) to determine the flux that would
fall into an aperture at a set distance from the source.  The
relationship between the flux and distance from the source was found
to be well fit by a semi-Gaussian profile with FWHM 
10\farcs5.  Flux from the target was $<$5\% of the
flux measured in an on-source aperture at $>$10\farcs9 and
$<$1\% of the on-source flux at $>$13\farcs5.  For
each of our target sources the area within 20\arcsec was checked for
detected sources in the 24$\mu$m image, and any possible contamination
calculated by multiplying the aperture flux by the value of the
Gaussian function described above at the appropriate distance.  In
order to account for sources that might be below the 3$\sigma$
detection limit in the 24$\mu$m image, but that could be contaminating
the target photometry, we also considered detections recorded by the
APEX module in the IRAC 8$\mu$m mosaic.  The relation between 8$\mu$m
and 24$\mu$m flux was calculated by comparing the values determined
for sources detected in both images.  The median and median absolute
deviation for the ratio of measured fluxes was $F_{8\mu m}/F_{24\mu m}
= 4.57\pm3.56$.  Any sources within 20\arcsec of our
target detected at 24$\mu$m had their contamination calculated, the
contaminating source was then removed from the list of detected
sources at 8$\mu$m and then the residual 8$\mu$m detection list was
checked for sources within 20\arcsec of the target.  The flux of these
sources was divided by 4.57 to get the 24$\mu$m flux and the
contamination in the target aperture calculated as above.  The median
and mean absolute deviation of the contamination of the targets is
7\% $\pm$ 6\% of the aperture flux.  Sources with
high levels of contamination ($>$13\%, $>$1$\sigma$ higher than
average) or with other issues that could affect
their photometry (e.g. high number of bad pixels, close to the edge of
the mosaic) are noted in Table \ref{tab:results}.  PRF photometry
should mitigate against the effects of source contamination through
the active deblend algorithm, and so we do not include a contamination
correction for sources with PRF photometry.  We used the above steps
to determine the levels of contamination that would have been seen in
aperture photometry for those sources with PRF photometry.  This was
used as a method of checking that there could be no sources in the
IRAC 8$\mu$m mosaic that might have been blended with the target in
the lower resolution 24$\mu$m mosaic.  In all cases the possible
contamination was $\le1$\% of the PRF photometry. 

After subtracting the contaminating flux from the MIPS 24$\mu$m photometry
an aperture correction of 0.433 magnitudes determined from bright
targets was applied.  A colour-correction was applied for each target
by interpolation from values tabulated in the MIPS data handbook.  We
used the zero-point of 7.14Jy listed in the handbook to convert the
fluxes into magnitudes. Photometric errors for MIPS 24$\mu$m
observations is 4\% \citep{engelbracht}.  We added this in quadrature
to statistical errors returned from the aperture/PRF photometry APEX
modules arising from pixel to pixel variations to give a final error
on the flux.  The final MIPS photometry is listed in Tables
\ref{tab:resultsbright} and \ref{tab:results}. Of the 132 targets in
the sample, 2 fall outside the 24$\mu$m mosaic (2\_195 and
2\_277). These targets also fall outside the field of view in the IRAC
images, and therefore they are not included in the table.   

\section{Detection of excess emission}

\subsection{Near-infrared excess}

The sensitivity of the IRAC data allows the detection of photospheric
emission down to the levels of photospheric emission of M dwarfs at
the distance of NGC 1960 ($\sim$1300pc).  Thus all the targets which
fall within the IRAC fields of view are detected with good
signal-to-noise ($>$5).  

\begin{figure*}
\begin{minipage}{8cm}
\includegraphics[width=8cm]{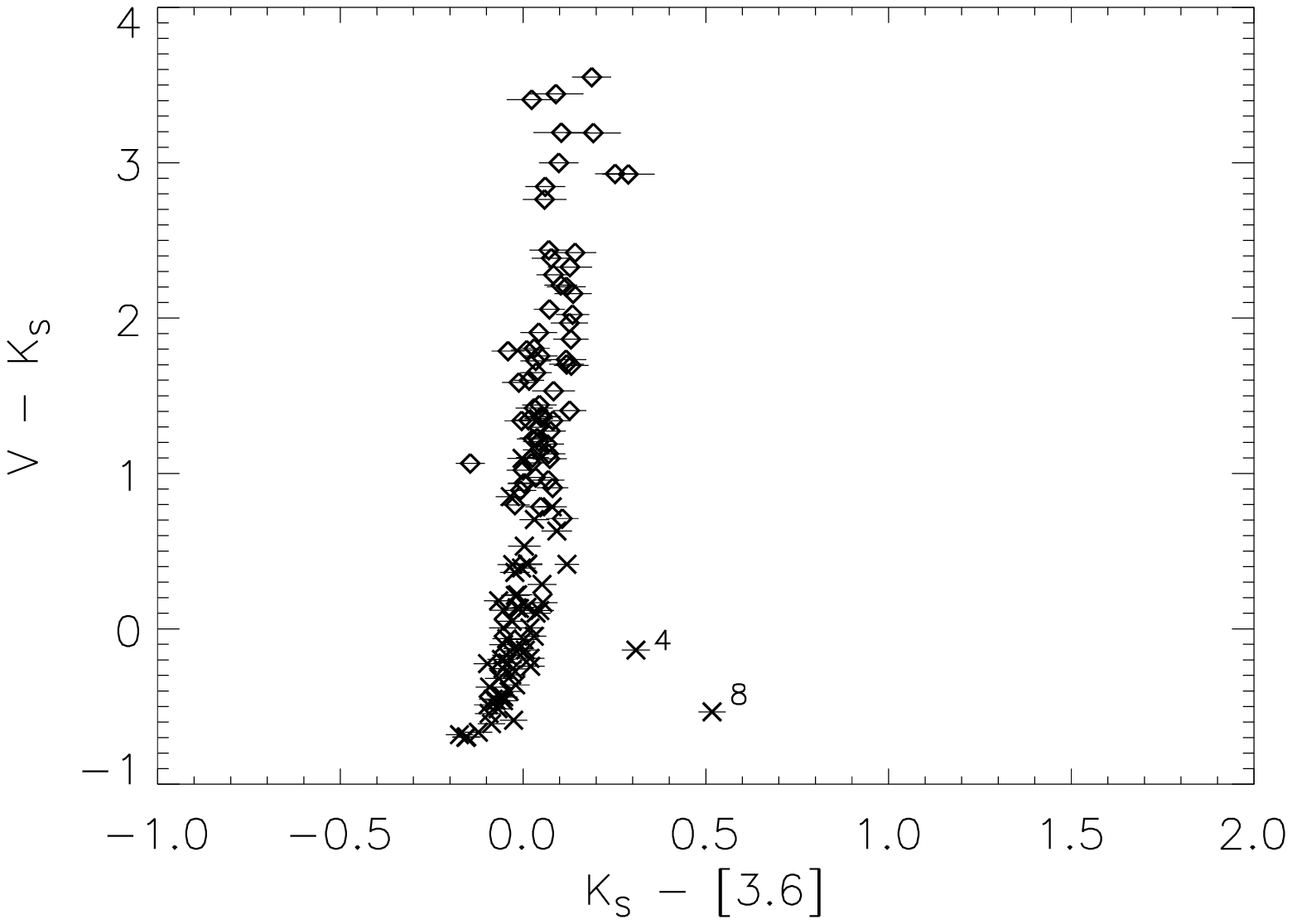}
\end{minipage}
\begin{minipage}{8cm}
\includegraphics[width=8cm]{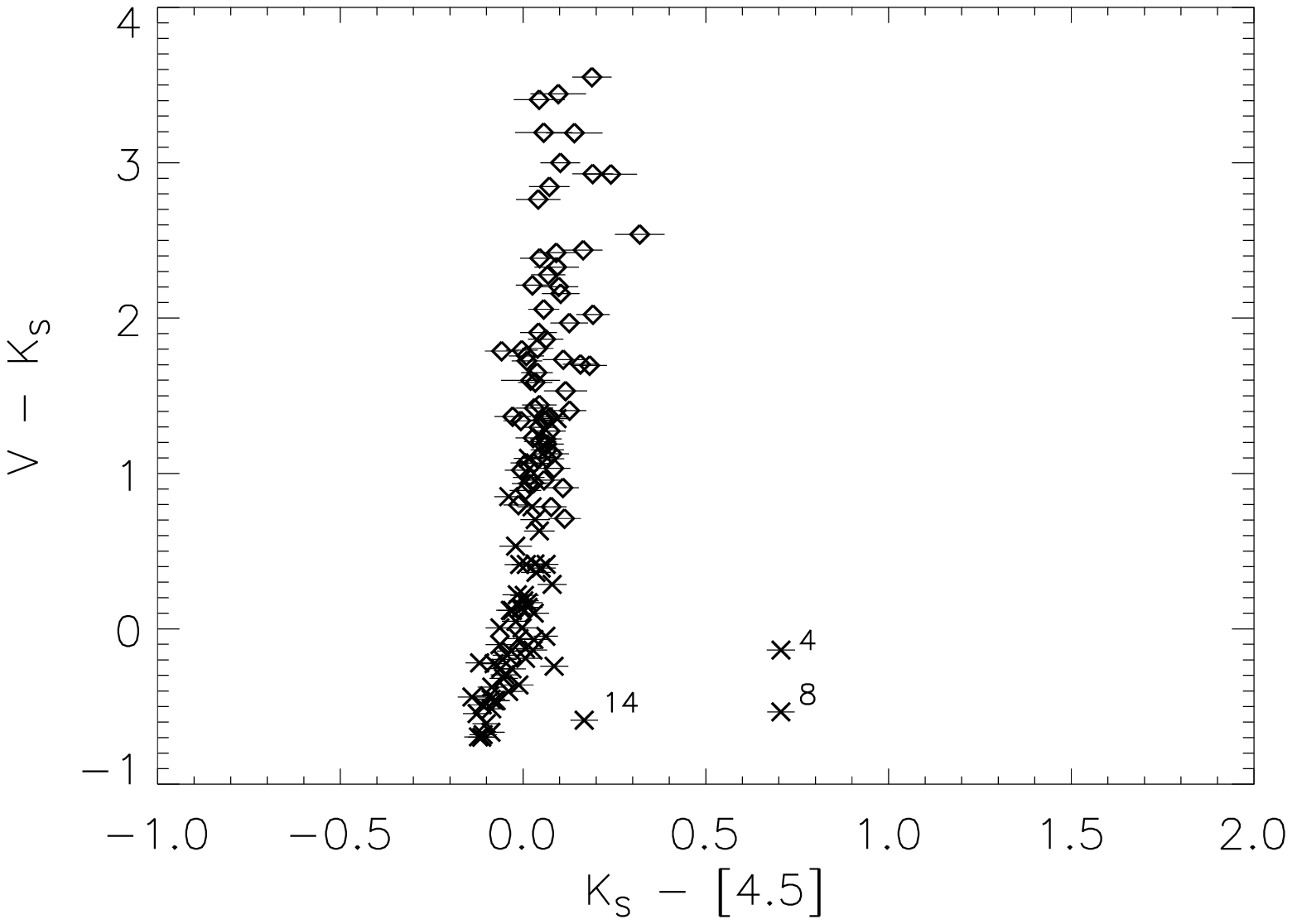}
\end{minipage} \\
\begin{minipage}{8cm}
\includegraphics[width=8cm]{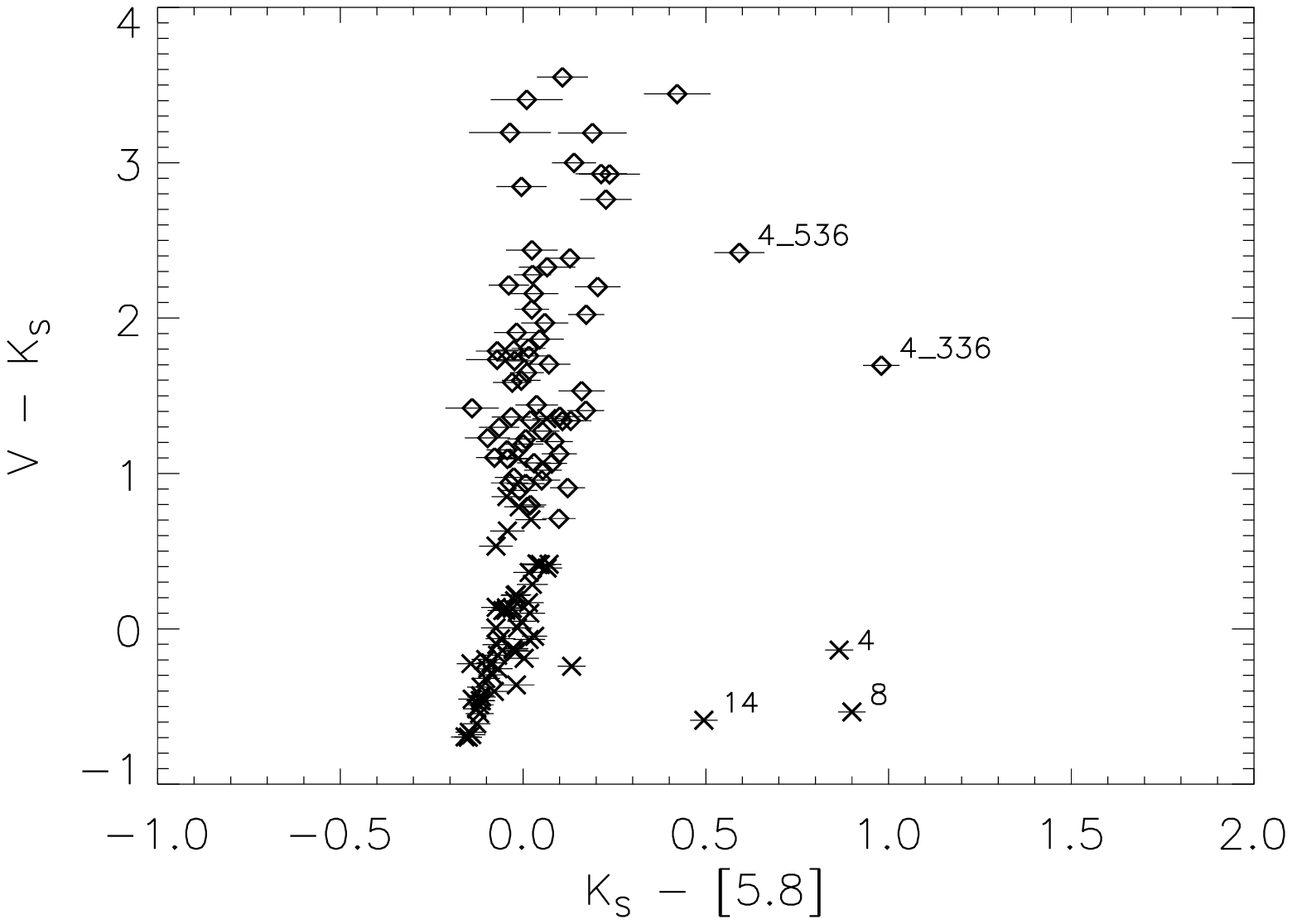}
\end{minipage}
\begin{minipage}{8cm}
\includegraphics[width=8cm]{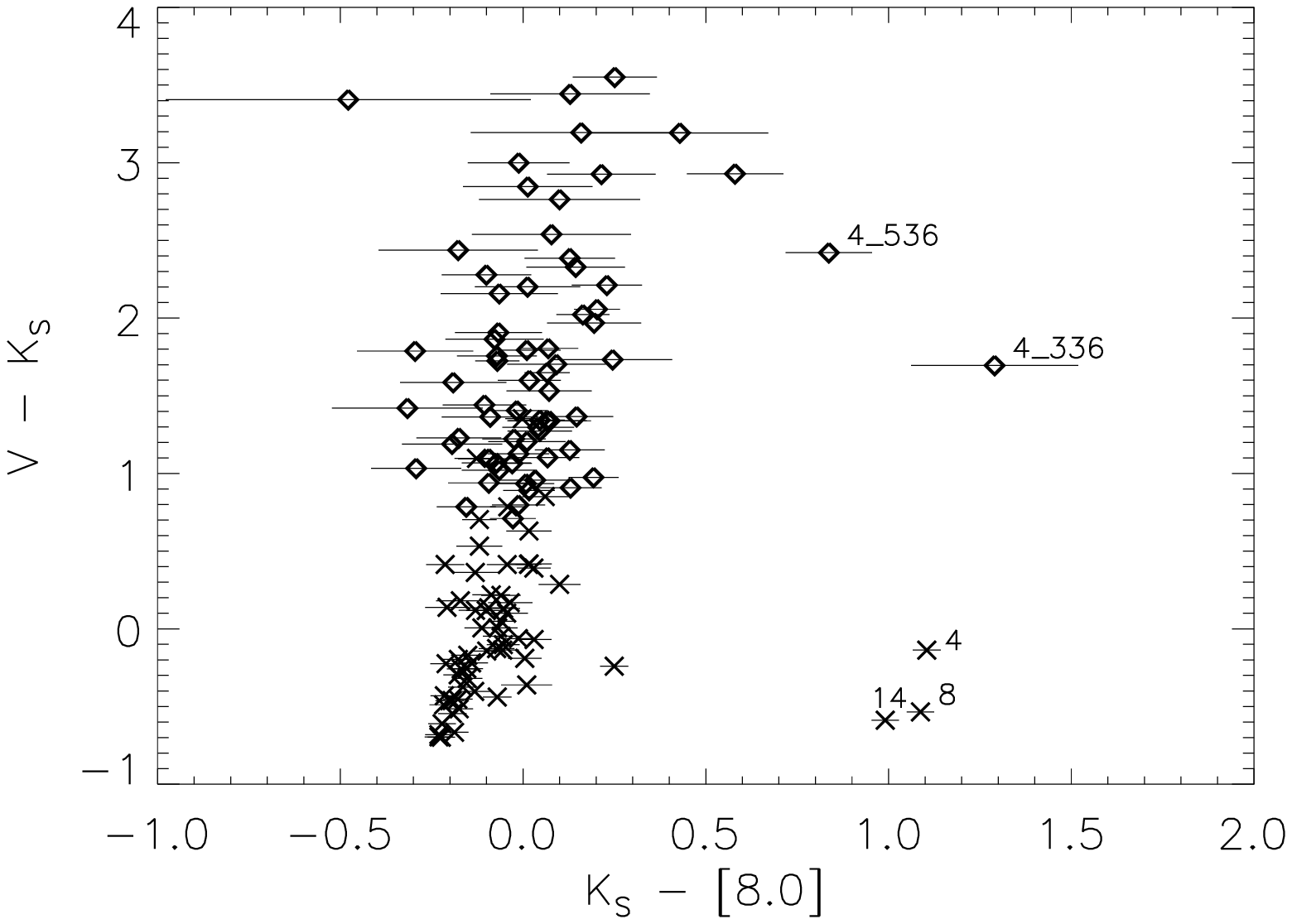}
\end{minipage}
\caption{\label{fig:iraccolours} The IRAC colour-colour plots of the
  targets in the cluster NGC 1960.   Errors include pixel-pixel
  variation and a 3\% calibration error on the IRAC data, added in
  quadrature to the uncertainty on the 2MASS $K_S$ band photometry.
  These plots show colours after dereddening (see text for
  details). Sources with colours indicative of a possible excess are
  labelled and discussed further in the text.} 
\end{figure*}

To determine if there is any excess emission in the IRAC photometry
from the target stars we plot the $K_S -[3.6]$, $K_S - [4.5]$, $K_S -
[5.8]$ and $K_S - [8.0]$ colours against $V - K_S$.  This method has
been used in other clusters to search for near-infrared excess
(see e.g. \citealt{gorlova_ngc2547, currie_persei, smith_ic4665}).  These
plots are shown in Figure \ref{fig:iraccolours}.  

The plots show that the majority of targets have colours consistent
with photospheric emission.  Crosses mark the targets in Table
\ref{tab:bright} and diamonds mark the sources listed in
Table \ref{tab:sources}.  Errors on the $K_S - $IRAC colours are taken from
the pixel-pixel variation in the background apertures in the IRAC
photometry plus the 3\% maximum calibration error typical for IRAC
photometry \citep{reach2005}, added in quadrature to the
errors on the $K_S$ photometry from 2MASS.  In order to identify
sources with unusual colours, we adopt the following procedure.  For
each target, we identify other targets in the sample with similar
$V-K_S$ colours (within $\delta(V-K_S) < 0.25$).  The
mean $K_S - $IRAC colour for these 
targets is calculated, together with the error (standard deviation of
the $K_S - $IRAC colours added in quadrature to the mean error on the
colour calculated as described above).  Any targets with colours
differing from the mean of other targets with similar $V - K_S$ by
more than 3 times this error are identified and removed from the source
list.  This process is repeated until no sources with colours
different from the average by more than 3 times the error are found.
The outlier sources are labelled on the figures.

From the IRAC colours, we identify bright targets 4 and 8 as having
colours consistent with excess emission in all 4 IRAC bands, bright
target 14 as having colours consistent with excess emission in bands
[4.5], [5.8] and [8.0], and lower mass stars 4\_336 and 4\_536 as
having colours consistent with excess in the two longest IRAC
wavebands.  None of these targets are highlighted in the 2MASS
catalogue as having uncertain $K_S$ photometry.  

Bright target 16 has colours that appear to be consistent with excess
emission in the two longest IRAC bands ($K_S - [5.8] = 0.133\pm0.038$
and $K_s-[8.0]=0.249\pm0.039$), although in the $\sigma$-clipping
algorithm described above these colours are at the 2.3$\sigma$ and
2.1$\sigma$ levels of significance.  Constructing an SED for this
object using a Kurucz model profile for the appropriate temperature
(9900K, see Table \ref{tab:bright}) and scaling to a best fit for the
2MASS $JHK_S$ photometry we find the IRAC channel 3 and 4 photometry
to be consistent with photospheric emission within 2 and 3$\sigma$
respectively.  Fitting a blackbody to the IRAC and MIPS photometry of
the source shows that any excess must be at a maximum temperature of
600K (in order to not exceed limits from the IRAC photometry).  This
corresponds to a minimum radial location of 0.67au assuming the
emitting grains behave like blackbodies.  However, as the IRAC
photometry is consistent with a photospheric emission within the
errors we cannot confirm that this target exhibits true near-infrared
excess.  

\subsubsection{Contamination by other sources} 

The target 4\_536 lies very close to the bright target 8.  In the IRAC
channel 4 data (8$\mu$m), the emission from bright target 8 extends to
the location of 4\_536, and so the target may have contaminated flux.
This is also true in the IRAC channel 3 image, although the emission
from bright target 8 is less extended in this image.  As the
resolution is higher for the lower wavelength channels, this is not
the case in the channel 1 and channel 2 images.  Thus we cannot
confirm excess for this target in the IRAC bands.  This target is not
detected at the 3$\sigma$ level in the MIPS 24$\mu$m imaging, and so
we have no other photometric measurements to further test whether
this target exhibits true excess emission.   

The target 4\_336  lies very close to a bright infrared source 
(detected in the IRAC images at 84.02458 degrees in RA,
34.103 degrees in Dec).  This source has been identified as IRAS
05327+3404, a probable pre-main sequence K2 class I/II star with a
large optically thick circumstellar disc producing a large infrared
excess (not thought to be related to NGC 1960, \citealt{magnier}). In
the high resolution data (IRAC channels 1 and 2) this source can be
seen as separate from the target 4\_336. However, in the IRAC channels
3 and 4 and the MIPS 24$\mu$m images, the bright source IRAS
05327+3404 extends to cover the location of the target 4\_336.  We can
therefore only provide upper limits to the flux for this target in
IRAC channels 3 and 4, and in the MIPS 24$\mu$m photometry.  

\subsubsection{True excess}

\begin{figure*}
\begin{minipage}{8cm}
\includegraphics[width=8cm]{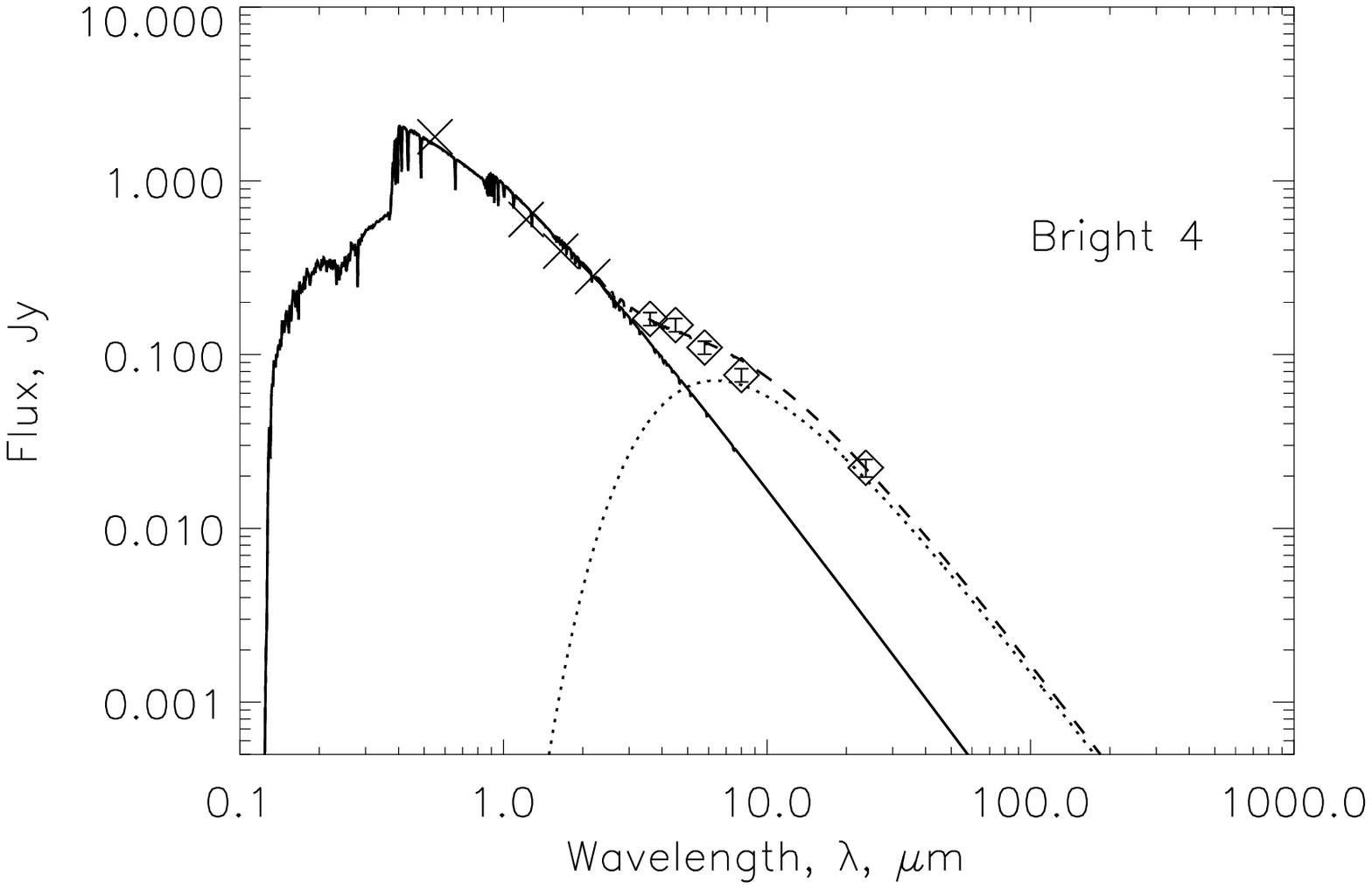}
\end{minipage}
\begin{minipage}{8cm}
\includegraphics[width=8cm]{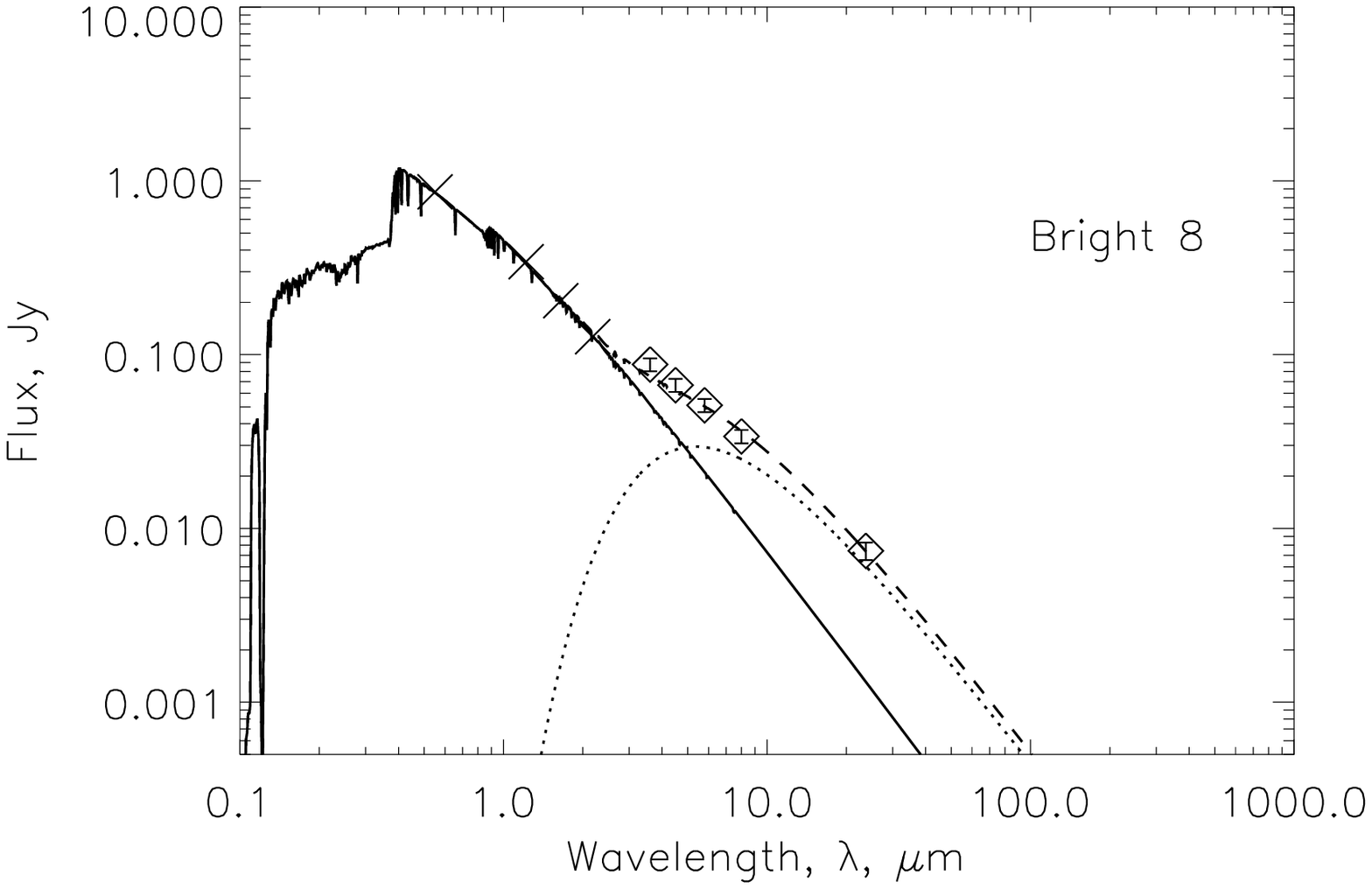}
\end{minipage} \\
\begin{minipage}{8cm}
\includegraphics[width=8cm]{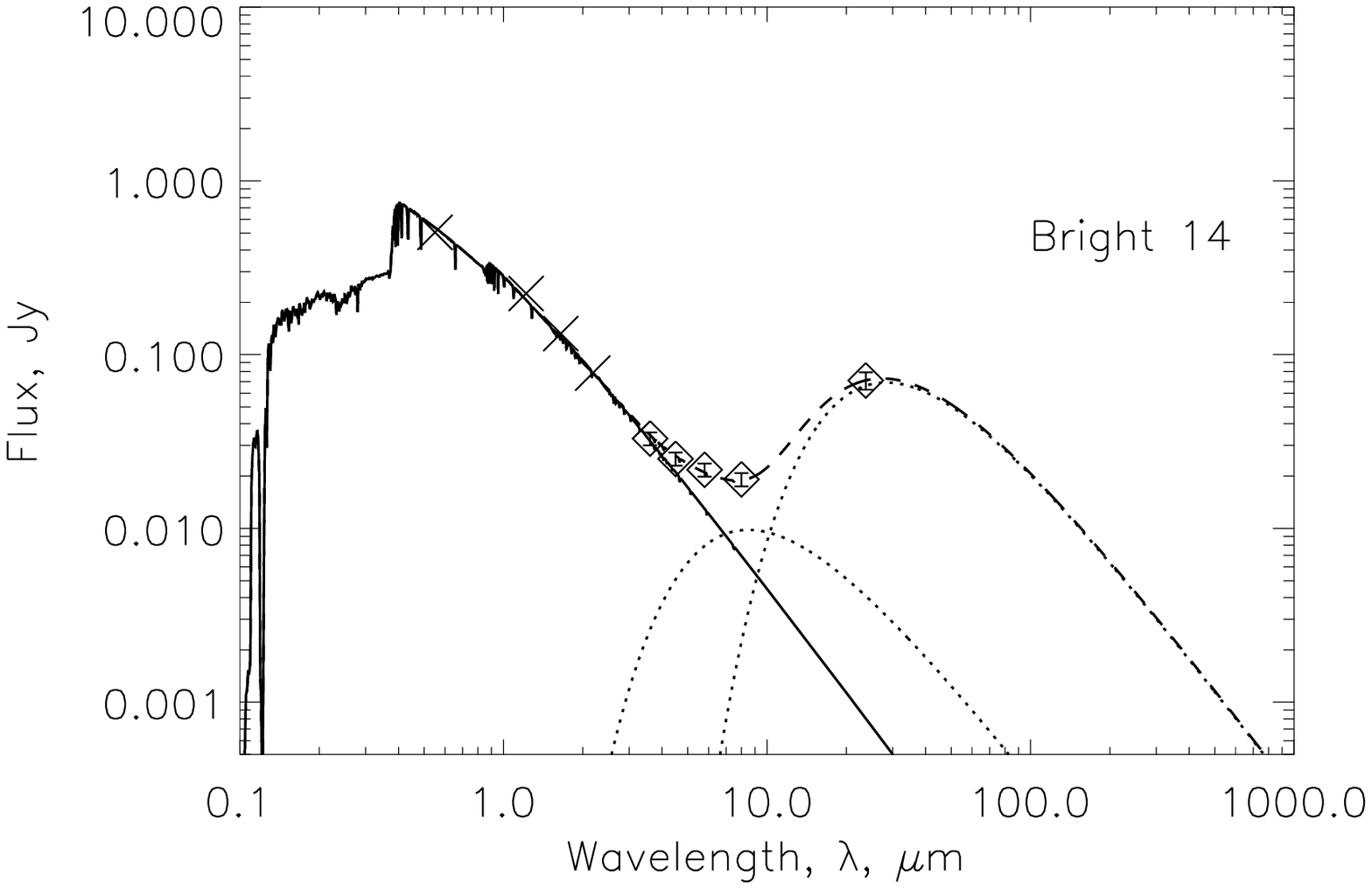}
\end{minipage}
\begin{minipage}{8cm}
\caption{\label{fig:nearIRxs} The SED of bright targets with
  significant excess emission in IRAC channels.  Crosses mark
  Hipparcos V band and 2MASS JHK$_{S}$ photometry.  Diamonds mark IRAC
  and MIPS 24$\mu$m photometry.  A Kurucz profile of appropriate
  temperature for each source (see Table \ref{tab:bright}) is shown
  with a solid line.  Dotted lines mark blackbody fits to the excess
  emission, with the total (blackbody plus photospheric) emission
  indicated by a dashed line.  }
\end{minipage} 
\end{figure*}

The bright targets 4, 8 and 14 (BD +34 1113, BD +34 1110, and BD +34
1098 respectively) are amongst the highest mass stars in
our sample, as indicated by their colour in $V-K_S$ and temperature
inferred from $B-V$.  These sources show evidence for excess emission
in the near and mid-infrared (see Table \ref{tab:resultsbright} and
Figure \ref{fig:iraccolours}).  We checked these sources did not have
significant contamination using the method described in section 2.
We construct SEDs for these objects 
using Kurucz model profiles for the appropriate temperature for each
source (see Table \ref{tab:bright}) scaled to a best fit to the
JHK$_{S}$ 2MASS photometry using a $\chi^2$ analysis.  These plots
(Figure \ref{fig:nearIRxs}) show that the near-infrared slopes are not
consistent with photospheric emission alone.  We can fit the IRAC and
MIPS photometry by adding a single temperature blackbody to the
photospheric emission for bright sources 4 and 8.  These fits are at
temperatures of 800K and 900K respectively (equivalent to dust lying
at radii of 0.73au and 0.36au respectively assuming the material emits
as a blackbody).  For bright target 14 the IRAC and MIPS photometry
cannot be fit by a single temperature blackbody, and so we adopt two
blackbody contributions at temperatures of 600K and 180K to fit the
emission (0.73 and 8.06au).  It should be noted that this illustrative
fit is non-unique and that longer wavelength photometry would be
required to determine constraints on the cooler blackbody temperature
in particular.  In all three plots the blackbody contributions are
shown by a dotted line and the total (blackbody and photospheric
emission) is shown by a dashed line (Figure \ref{fig:nearIRxs}).

For all three targets with near-infrared excess we must ask if this
emission is evidence for a remnant primordial disc.  At an age of
16Myr for the cluster such primordial discs would be rare as they are
expected to dissipate on timescales of a few Myr (see
\citealt{wyattreview} and references therein).  The
fractional luminosity of these discs is much lower than we would
expect for primordial disc emission.  Adopting the blackbody fits to
the emission shown in Figure \ref{fig:nearIRxs}, we find that the
excess emission has fractional luminosity $L_{\rm{d}}/L_\star =
3.9\times10^{-3}$ for both bright star 4 and 8, and that the two
components of the excess around bright star 14 have fractional
luminosities of $L_{\rm{d}}/L_\star = 1.3\times10^{-3}$ and
  $2.7\times10^{-3}$ (for components at 600K and 180K respectively).
  These levels are consistent with $L_{\rm{d}}/L_\star < 10^{-2}$
  which conventionally defines a debris disc as opposed to a
  primordial disc \citep{lagrange}.  These targets join a small but
  growing population of A stars with debris discs within 10au
  (e.g. $\zeta$ Lep, \citealt{moerchen}, HD172555 \citealt{rebull}).
  Bright star 14 is one of a growing number of systems with evidence
  for multiple component discs.  11 stars were identified by
  \citet{chen09} as having probable multiple component discs, which
  have been interpreted as possible solar-system analogues (see also
  the $\beta$ Pic association member $\eta$ Tel, \citealt{smitheta}).

\subsubsection{Limits on near-infrared excess and primordial emission} 

\begin{figure}
\includegraphics[width=8cm]{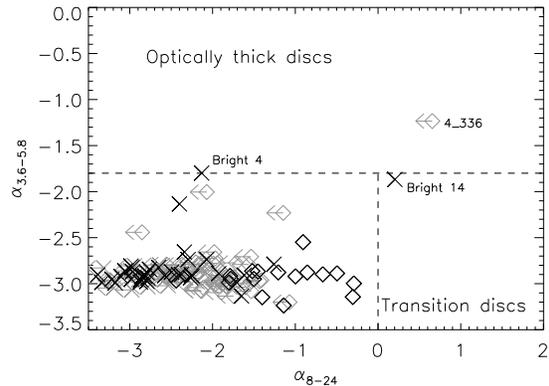}
\caption{\label{fig:transition} The spectral slopes for all targets in
  our study.  Sources without significant detections at 24$\mu$m are
  shown as upper limits to the spectral slope based on the 3$\sigma$
  upper limits to their flux.  Regions of the spectral slope
  parameter space which could indicate primordial (optically thick) or
  transition discs are labelled.  Most targets have spectral slopes
  consistent with no excess or optically thin emission consistent with
  a debris disc.  See text for discussion.}
\end{figure}

As a further confirmation of the evolved nature of the targets
observed in this study, we follow the example of \citet{muzerolle10}
and consider the IRAC/MIPS SED slope for each of our objects as a
method of identifying possible primordial/transition disc candidates.  
We consider the value of $\alpha = d \log{\nu F} / d \log{\lambda}$
over IRAC channels 1 and 3 (so for $\lambda = 3.6\mu$m and $5.8\mu$m)
and over IRAC channel 4 and MIPS ($\lambda = 8.0\mu$m and $23.7\mu$m).
We plot these spectral slopes for all stars in our study in Figure
\ref{fig:transition}, and plot upper limits to the spectral slope
$\alpha_{8-24}$ for sources without significant photometry at
$24\mu$m.  According to the metrics of \citet{muzerolle10}, stars with
$\alpha_{3.6-5.8}>-1.8$ are possible primordial disc hosts.  Stars
with $\alpha_{3.6-5.8}<-1.8$ and $\alpha_{8-24}>0$ are possible
transition disc candidates, and may have optically thick discs at
$>$1au with depleted/optically thin inner holes.  Target 4\_336
appears to lie in the primordial disc region, but this object's flux
is highly contaminated by a known primordial disc candidate as
discussed in Section 4.1.1, and therefore this is not a true detection
of a primordial disc.  Bright target 4 lies
close to the boundary for possible primordial emission.  Bright target
14 lies in the region of possible transition disc emission.  However,
for both these targets we have shown that it is possible to fit their
emission with optically thin ($L_{\rm{IR}}/L_\star < 10^{-2}$)
blackbody emission.  Thus we find the emission for both these targets
is consistent with debris disc emission.  The remaining targets in our
study all have $\alpha_{3.6-5.8}<-1.8$ and $\alpha_{8-24}<0$,
consistent with no excess or optically thin emission.  

\begin{figure*}
\includegraphics[width=12cm]{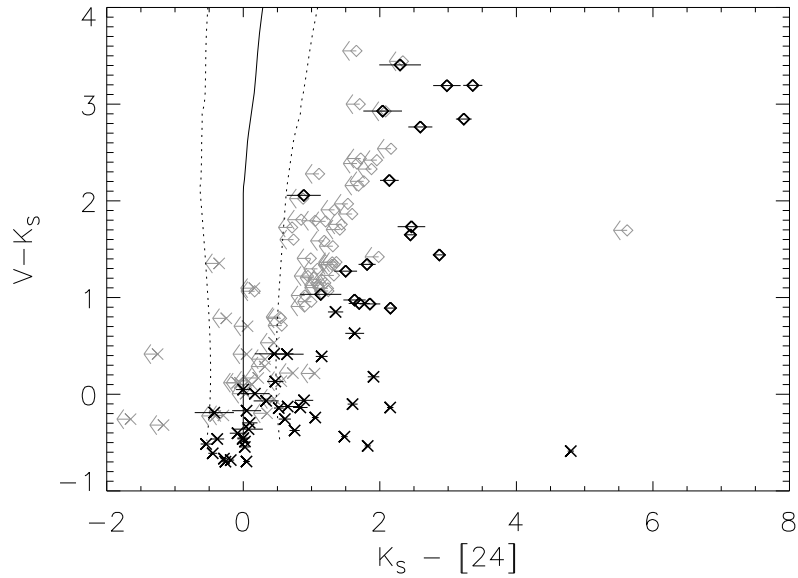}
\caption{\label{fig:mips} The $K_S - [24]$ vs $V - K_S$ colour-colour
  diagram for sources with $> 3\sigma$ detections in the MIPS mosaic.
  Grey symbols mark upper limits for sources with $< 3\sigma$
  detections.  Diamonds mark sources from the lower mass 
  sample, crosses mark additional sources from \citet{sanner}.  Errors
  are the statistical and calibration error on [24] added in
  quadrature to the error on the 2MASS $K_S$ magnitude.  The
  predicted photospheric colours from \citet{plavchan} are shown with
  a solid black line.  For most of the spectral range under
  consideration, only sources with significant excess are detected.}
\end{figure*}

For those targets with no evidence of significant near-infrared
excess, we determine the limits we can place on excess emission in the
IRAC photometric bands.  Excluding the sources discussed above, we
determine an empirical photospheric locus and uncertainty in $K_S -
IRAC$ as a function of $V-K_S$.  For each value of $V-K_S$, we
determine the mean and standard deviation of the $K_S - IRAC$ colours
of sources with $V-K_S$ within 0.25 of this value.  These are our
empirical photospheric locus and error.  We fit a linear model to the
error to determine the sigma values for each colour.  We determine the
dispersions in $K_s - IRAC$ colours to be $\sigma_{K_S-[3.6]} =
0.04+0.005(V-K_S)$, $\sigma_{K_S-[4.5]} = 0.04+0.01(V-K_S)$,
$\sigma_{K_S-[5.8]} = 0.04 + 0.02(V-K_S)$, and $\sigma_{K_S-[8.0]} =
0.05+0.04(V-K_S)$.  Adopting these uncertainties, we find that for an
F5-type star ($V-K_S \sim 1$) we can place the following 3$\sigma$
limits on excess emission in the near-infrared: $F_{3.6}/F_{\rm{phot}}
< 1.13$; $F_{4.5}/F_{\rm{phot}} < 1.15$; $F_{5.8}/F_{\rm{phot}} < 1.18$; 
and $F_{8.0}/F_{\rm{phot}} < 1.28$.  For objects with $V-K_S = 4$
($\sim$ M1-type) we place the following 3$\sigma$ limits on excess in
the near-infrared: $F_{3.6}/F_{\rm{phot}} < 1.18$;
$F_{4.5}/F_{\rm{phot}} < 1.25$; $F_{5.8}/F_{\rm{phot}} < 1.39$;  and
$F_{8.0}/F_{\rm{phot}} < 1.79$.

\subsection{24$\mu$m excess}

The principal means of identifying debris discs is the detection of
excess emission (above the levels expected from the photosphere) in
mid-infrared ranges, and no detection of such an excess at
near-infrared wavelengths (which may indicate a primordial or
transition disc).  The sensitivity of the MIPS observations of NGC 1960
allow a detection down to $\sim$12th magnitude at 24$\mu$m (3$\sigma$
point source limit).  This means we can only detect photospheres to A3V
($V-K < 0.10$), which means that for many of our targets, we would not
expect to detect the photosphere.  This is reflected in the results
presented in Tables \ref{tab:resultsbright} and \ref{tab:results}.
Those sources with significant detections generally have high values
of $K_S - [24]$ and $F_{24}/F_{\rm{phot}}$ (detected emission at
24$\mu$m / expected emission from the photosphere).  

In order to determine which targets exhibit 24$\mu$m excess, we follow
the example of recent authors (e.g. \citealt{rebull, stauffer}) in
using the targets' $K_S - [24]$ colours. This method requires a
well-defined model for photospheric colours.  \citet{stauffer} used
Spitzer observations to determine an empirical relation for $K_S -
[24]$ from $V-K_S$, defined as $K_S - [24] = 0.042-0.053\times(V-K_S)
+ 0.023\times(V-K_S)^2$.  Similar relations have been presented by
\citet{gorlova}, and \citet{plavchan}, who included M dwarfs in the
spectral range covered.  As the \citet{stauffer} relation is only
valid to $V-K_S < 3$ and our NGC 1960 sample includes sources of lower
mass, we use the \citet{plavchan} relation as our photospheric model.  

\begin{figure*}
\begin{minipage}{8cm}
\includegraphics[width=8cm]{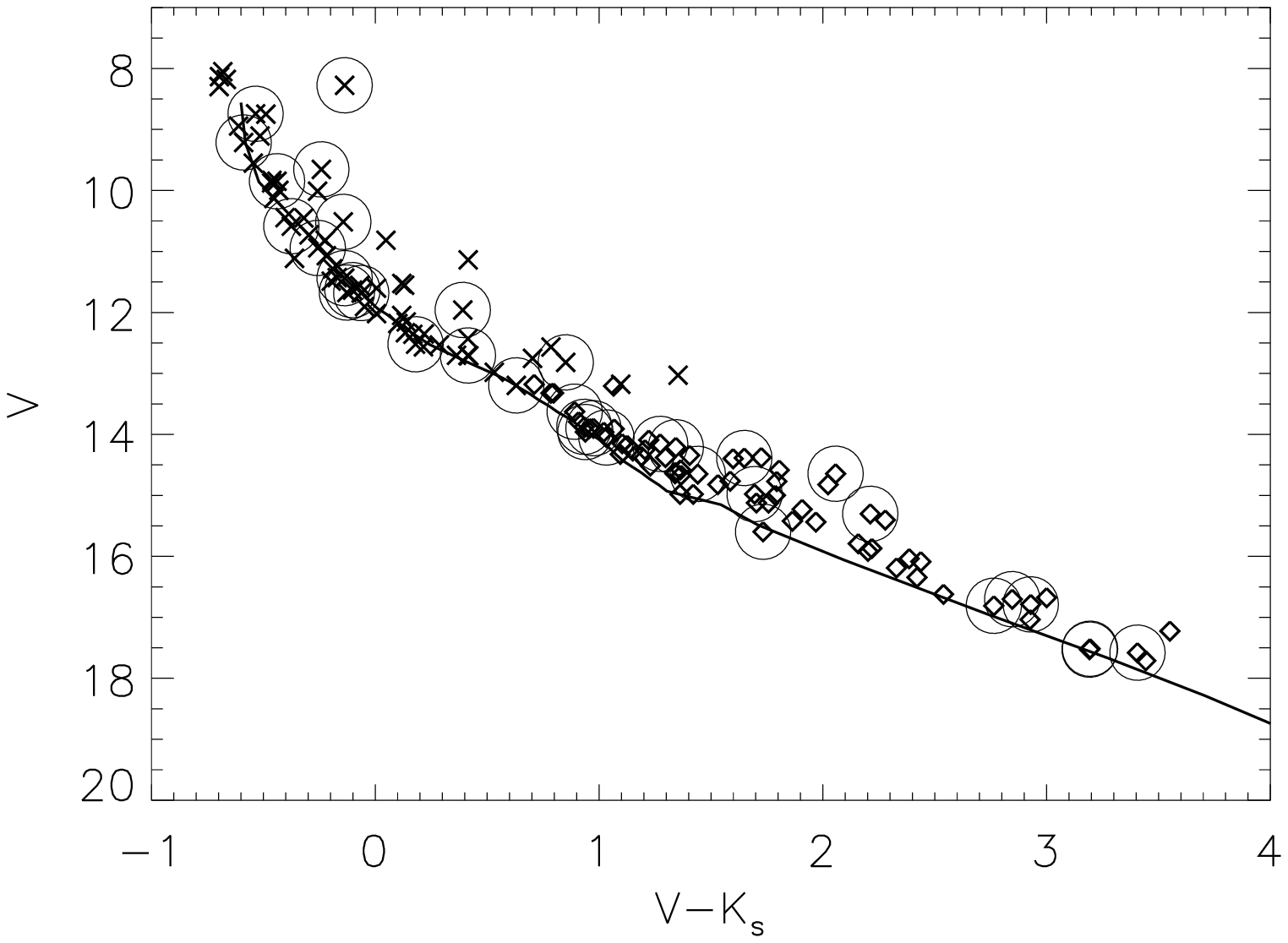}
\end{minipage}
\begin{minipage}{8cm}
\includegraphics[width=8cm]{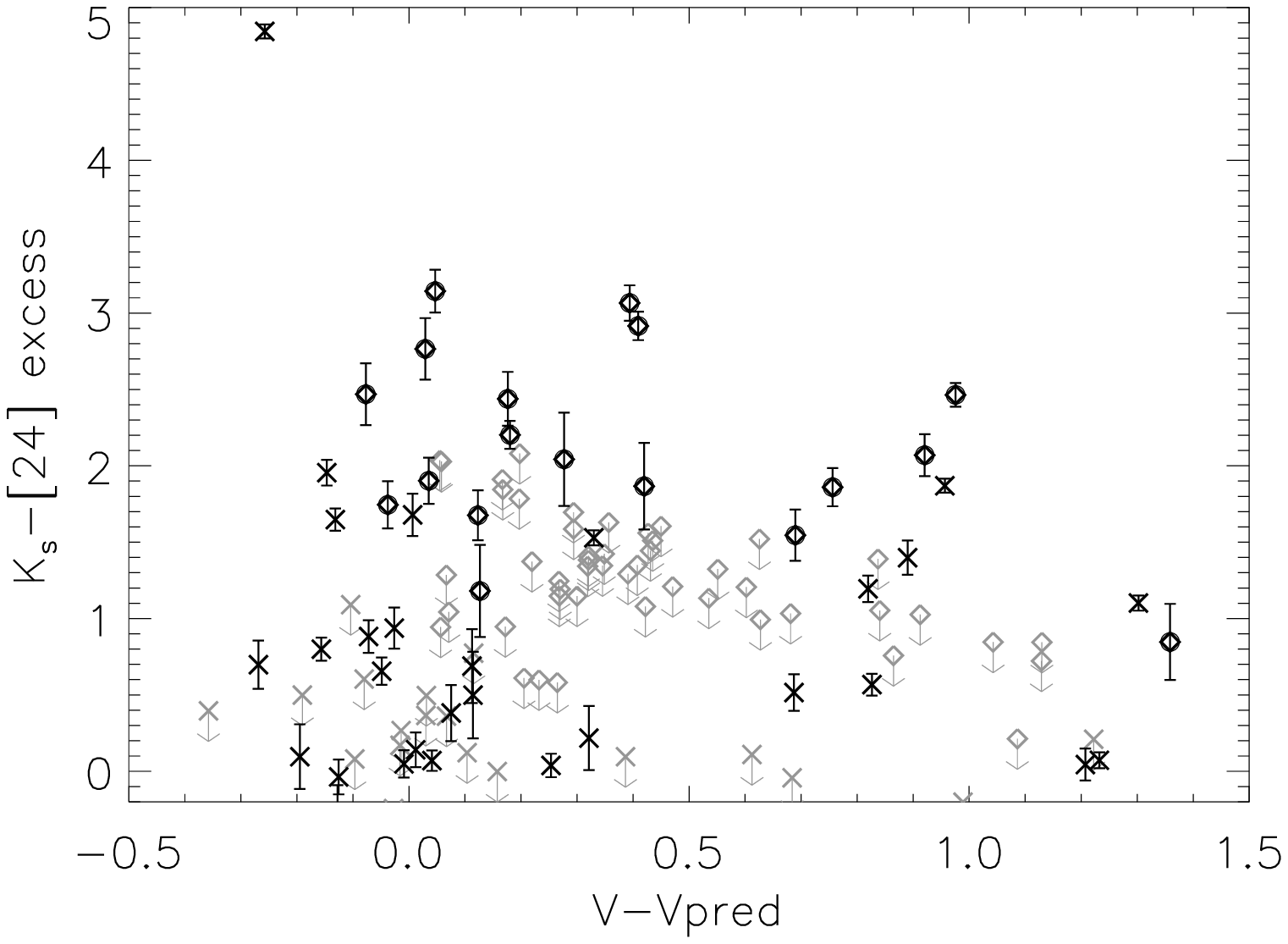}
\end{minipage} \\
\caption{\label{fig:binarity} Dependence of excess on a proxy for
  stellar multiplicity.  \emph{Left:} Height above a single star
  isochrone of 16Myrs is used following \citet{stauffer}.  Diamonds mark the
  targets listed in Table \ref{tab:sources}, and crosses the targets
  in Table \ref{tab:bright}.  Stars identified as having excess
  emission (see Tables \ref{tab:resultsbright} and \ref{tab:results})
  are overplotted with large circles.  Magnitudes and colours
  are corrected for reddening and extinction.  \emph{Right:} Height above the
  isochrone is compared with $K_S - [24]$ colour.  There is no
  evidence of a significantly different distribution of colours for
  suspected multiples ($V-V_{\rm{pred}} > 0.3$). }
\end{figure*}

In Figure \ref{fig:mips} we show the $K_S - [24]$ vs $V - K_S$ colours
of the target sources.  Crosses are used to mark the bright
targets taken from \citet{sanner}, and diamonds mark the sources
identified as lower mass cluster members.  Upper
limits for targets with signal to noise of less than 5 are marked in
grey.  Over-plotted as a solid line is the expected photospheric
colour from \citet{plavchan}.  We determine the dispersion around this
relation by considering the colours of the bright targets
($V-K_s<0.1$) for which the photosphere should have been detectable.
A histogram of the $K_S - [24]$ colours of these sources was
constructed, and a Gaussian function fitted to the left-hand side of
this function (the targets which are not likely to have excess
emission).  The best fitting Gaussian FWHM was found to be 0.14.
This dispersion includes the errors in $K_S -[24]$ (typically around
0.05 for targets with $V-K_S < 0.1$, see Table
\ref{tab:resultsbright}), and so we adopt this value as a minimum
uncertainty on the photospheric values of $K_S - [24]$. We
expect the errors on the $K_S - [24]$ colours to
increase for fainter targets due to decreased signal to noise.  Using
all significant detections ($>5\sigma$) we examined the errors on the
24$\mu$m photometry as a function of $V-K_S$.  We found that the
effect of source colour on the uncertainty was relatively weak.  This is
because for most of the spectral range considered in this study we are
detecting sources with excess emission only, and so the error levels
were nearly constant.  We determined a linear fit to the uncertainty (using
a $\chi^2$ test) on $K_S - [24]$ and added this to the measurement of
the dispersion adopted as a minimum uncertainty as discussed above.
This gave us a final uncertainty 
on $K_S - [24]$ from our photometry of $0.14 + 0.032(V-K_S)$.
We also consider the differences between the photospheric colour
models from \citet{plavchan}, \citet{gorlova} and \citet{stauffer}.
For each value of $V-K_S$ we interpolate the predicted $K_S-[24]$ for
each model, and add the standard deviation of the three results in
quadrature to the error relation already described.  These errors are
shown (at a 3$\sigma$ level) by a dotted line in Figure
\ref{fig:mips}. 

From this figure we identify 17 of the targets from Table
\ref{tab:bright}, and 18 of those from Table \ref{tab:sources} as
having significant 24$\mu$m excess.  These sources are identified in
Tables \ref{tab:resultsbright} and \ref{tab:results} by a note in the
Comments column.

\section{Discussion}
\subsection{Relationship with multiplicity}

As the majority of stars are binary or higher-order multiples
\citep{abt,duquennoy} , the
relationship between stellar multiplicity and debris disc parameters
has been the subject of much recent study.  \citet{cieza} presented  
an analysis of IRAC data from several Spitzer legacy surveys which
showed that for projected separations of $<$40au, systems were half as
likely to retain their primordial discs than systems with larger
separations (suggesting a disc lifetime of 0.3-0.5 Myr for close
binaries compared to 3-5 Myr around single stars).  Conversely
\citet{trilling} found in their study of field stars (with most stars
$>$600 Myr old) that the binary stars had a higher incidence of debris discs
than single stars.  Debris discs were found to be more common around
binaries with small ($<$3au) or large ($>$50au) separations than
around intermediate separation binaries (3--50au).  This survey
concentrated on detections at 70$\mu$m, 
and so on discs that were further from their host stars.
\citet{plavchan} found no evidence of the trend suggested by
\citet{trilling}, and  \citet{duchene} found no significant dependence
of debris disc incidence with binarity or binary separation. 
\citet{stauffer} showed tentative evidence that in the 100 Myr Blanco 1
cluster binary stars have a lower 24$\mu$m excess frequency than
single stars (adopting height above a single star isochrone as a 
binarity proxy).  They combine their results with data
from the Pleiades ($\sim$ 100 Myr) and NGC 2547 ($\sim$ 35 Myr) and find
an overall chance of 0.05\% that the excess around single and binary
samples is drawn from the same parent population (using a K-S test).
In \citet{smith_ic4665} we showed that there was no evidence for a
dependence of debris disc emission on stellar multiplicity in the
27 Myr old open cluster IC4665.  We adopt the same analysis here using
height above a single star isochrone as a proxy for multiplicity and
compare this to $K_S - [24]$ colour.  The isochrone was tuned from a
fit to the Pleiades following \citet{stauffer}.  This is shown in
the left-hand panel of Figure \ref{fig:binarity}. Bright star 4 (HIP
26354) stands out in this plot as having a very high $V$ band
magnitude for its colour ($V-K_S$).  This source has been identified
as having high variability ($\delta V=0.281$ \citealt{lefevre}) and
therefore the target's photometric measurements are dependent on the
epoch of observation, a possible cause of the anomalously high $V$
band magnitude.

We expect that most targets will have $V - V_{\rm{pred}} \leq 0.75$
(where $V_{\rm{pred}}$) is the predicted V band magnitude of a star of
a given $V-K_S$ colour from the single star isochrone). Sources with
$V-V_{\rm{pred}} > 0.75$ are possible triples or higher-order systems.
From our sample of 132 targets (from Tables \ref{tab:sources} and
\ref{tab:bright}) 26 are found to have $V-V_{\rm{pred}} > 0.75$,
giving a detection rate of $19_{-5}^{+5}$\% higher-order multiples.
This is somewhat higher than is seen in other studies ($\sim$9--12\%,
\citealt{abt, abt&levy, duquennoy}), although not significantly so.
This could also be reduced if we take into account errors on the 2MASS
$K_S$ photometry, which means that at the 3$\sigma$ level a minimum of
16 sources (12\% of the target list) have $V - V_{\rm{pred}} > 0.75$,
which agrees with the expected frequencies of higher-order
multiplicity.  In general we do not see any evidence for a relation
between $V-V_{\rm{pred}}$ 
and $K_S - [24]$ (see right-hand panel of Figure \ref{fig:binarity}).
We separate the sample into sources with $V - V_{\rm{pred}} \le 0.3$
and those with $V - V_{\rm{pred}} > 0.3 $ and compare the $K_S - [24]$
colours of the two populations.  Using a K-S test on the $K_S - [24]$
excess for sources with significant 24$\mu$m detections (where the
excess is defined as $(K_S - [24])_{\rm{obs}} - (K_S -
[24])_{\rm{pred}}$ and $(K_S - [24])_{\rm{pred}}$ is determined from the
Plavchan model for the photospheric colours) we find only a 70\%
probability that we can reject the null hypothesis that the
samples are drawn from the same population.  We also compare the two
samples ($V-K_S \leq 0.3$ and $V-K_S > 0.3$) including the upper
limits.  We use the ASURV statistical package \citep{asurv} to compare
the two populations using several tests (see \citealt{feigelson} for a
description of the tests).  We find that the probability
of rejecting the null hypothesis (that the samples are drawn from the
same population) is 0.47 according to the logrank test, and 0.74 according to
the Peto-Prentice test.  The differences in the probabilities arise
from the different treatments of the upper limits in the methods.
In either case we see no evidence that the multiplicity of the stellar
system affects the incidence of debris disc emission in this cluster.

\subsection{Placing NGC 1960 into context}

The NGC 1960 cluster is at an interesting age for planet formation.
At 16Myr, any gas giant planets are expected to have formed and the
circumstellar disc evolved from a gas-rich to gas-poor state within
$\sim 3$Myr for high-mass stars (B, A and early F-type,
e.g. \citealt{hernandez05}) or 5--7Myr for low mass stars (F5 or
later, e.g. \citealt{hernandez}).  Terrestrial planet formation on
the other hand is expected to continue for up to 100Myr
(e.g. \citealt{weidenschilling}).  A peak in 24$\mu$m debris emission
at 10--20 Myr around A and early F-type stars can be interpreted as
evidence of delayed stirring of the disc by the formation of
Pluto-sized bodies at this epoch (see e.g. \citealt{kenyon02}). We now
explore how NGC 1960 fits in with these models.  

\subsubsection{Excess emission at 24$\mu$m} 

A `rise and fall' in 24$\mu$m excess emission (emission above expected
photospheric levels) was first identified by
\citet{hernandez} and \citet{currie_persei}.  The peak levels of
excess emission (defined as emission above that arising from the
stellar photosphere) and frequency of excess are seen to rise from
$\sim$5--10 Myrs, before falling after $\sim$ 20 Myrs around
higher mass (A and early F-type) stars.  We examine NGC 1960 in the
context of this pattern by comparing the rate of excess around the
higher mass stars to other clusters of ages $\ge$5 Myrs.  

\begin{figure}
\includegraphics[width=8cm]{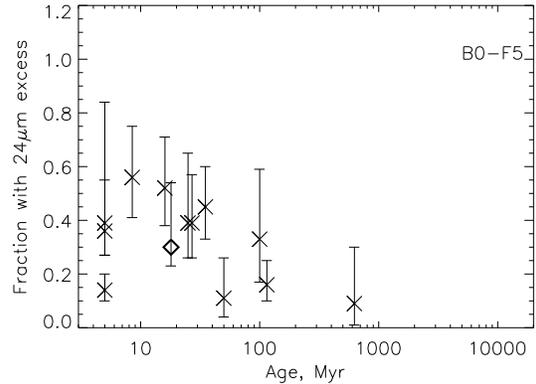}
\caption{\label{fig:astar_freq} The frequency of 24$\mu$m excess
  emission of stars of spectral types B0-F5 as a function of age.
  Clusters are listed in the text and in Table \ref{tab:xs_freq}.
  The datum for NGC 1960 is shown by a diamond and is offset in age to
  18 Myr to avoid confusion with data from the Sco-Cen association. } 
\end{figure}

In Figure \ref{fig:astar_freq} we show the frequency of excess
emission for stars of spectral types B0-F5 observed with MIPS.  
We include data for the following open clusters: $\lambda$ Orionis (5Myr,
\citealt{hernandez09,hernandez10}); 
Upper Scorpius (5 Myr, \citealt{carpenter} and \citealt{chen_scocen});
Orion OB1b (5 Myr) and Orion OB1a (8.5Myr, both \citealt{hernandez06}); 
Lower Centaurus Crux (16Myr) and  Upper
Centaurus Lupus (17Myr, both \citealt{chen_scocen}); NGC 2232 (25Myr,
\citealt{currie_ngc2232}, note that only 38/209 stars have published MIPS
24$\mu$m observations); IC 4665 (27Myr, \citealt{smith_ic4665});
NGC 2547 (35Myr, \citealt{gorlova_ngc2547}); IC 2391 (50Myr,
\citealt{siegler}); Blanco 1 (100Myr, \citealt{stauffer}); the
Pleiades (115Myr, \citealt{gorlova}); the Hyades (625Myr, \citealt{cieza});
and a collection of field stars \citep{bryden}.  For all clusters
we extract the spectral type for each object listed in the above
papers.  Where no spectral type is listed we assign a spectral type
based on the $V-K_{S}$ colour of the target.  The relationship was
determined from Kurucz model profiles of stars at typical temperatures
for each spectral type (so F5 stars have $V-K_S \sim 1$, temperatures
taken from \citealt{allen}).  The frequencies are
listed in Table \ref{tab:xs_freq}.  Clusters with less than 10 members
in the spectral range B0-F5 were excluded from Figure
\ref{fig:astar_freq} but are listed in Table \ref{tab:xs_freq} and
included in our examination of levels of excess emission below.  Stars
are identified as having excess emission if they satisfy the
following criteria (following \citealt{currie_ngc2232}):
\begin{equation}\label{eq:xs_sig} (K_S - [24])_{\rm{obs}} - (K_S -
  [24])_{\rm{phot}} \ge 3\sigma_{[24]} \end{equation} and 
\begin{equation} \label{eq:xs_phot} (K_S - [24])_{\rm{obs}} - (K_S -
  [24])_{\rm{phot}} \ge 0.15, \end{equation} where this last
condition takes into account systematic errors, e.g. from dispersion
in $K_S-[24]$ colours.  It should be noted that all the excess sources
identified in NGC 1960 meet these criteria.  $(K_S -
[24])_{\rm{phot}}$ for each target is   
given by the relation determined by \citet{plavchan}, as shown in
Figure \ref{fig:mips}.  For this spectral range we expect $K_S -
[24] \sim 0$ (dereddened).  Extinction ($A_V$) was taken from the
papers mentioned above, and converted to $A_{K_S}$ and $A_{24}$ according
to the relations in \citet{rieke&lebofsky} and \citet{flaherty}
respectively.  

In Figure \ref{fig:astar_freq} and Table \ref{tab:xs_freq} we give a
1$\sigma$ confidence limit on the frequency of excess.  We assign a
confidence limit based on the tabulated confidence limits of Poisson
statistics presented in \citet{gehrels}.  However, for several of the
samples listed in Table \ref{tab:xs_freq} there are sources with
24$\mu$m upper limits or non-detections, which could potentially be
disc hosts.  For all sources with upper limits to their 24$\mu$m
photometry, we take the 3$\sigma$ upper limit to [24] and determine if
the source could have an excess.  This is determined by checking if
the upper limit would be listed as an excess according to equations
\ref{eq:xs_sig} and \ref{eq:xs_phot}.  If the upper limit is below
these cut-offs, the source is an upper limit with no disc.  Otherwise
the target is a potential host of excess emission.  We use the number
of sources with possible excess emission to define another upper limit
to the confidence limits on the frequency of excess for each cluster.
The largest of this upper limit and that determined from the Poisson
statistics of \citet{gehrels} is presented in Table \ref{tab:xs_freq}
and shown in Figure \ref{fig:astar_freq}.  In
this plot we can see that the cluster NGC 1960 has a fairly typical
frequency of excess emission given its age ($30^{+24}_{-7}$\%).  We
can possibly see the peak of 
excess frequency at $\sim$10--20 Myr, however the statistics for the
younger clusters are quite uncertain.  As only one young cluster
(Upper Sco, $\sim$5 Myr) has a low frequency of excess, we cannot
confirm with this data that the ``rise and fall'' in frequency is
statistically significant.

\begin{table*}
\caption{\label{tab:xs_freq} The excess frequencies of open clusters
  with published 24$\mu$m MIPS photometry.}
\begin{tabular}{lrrcrcl}\hline Name & Age & \multicolumn{2}{c}{B0-F5}
  & \multicolumn{2}{c}{F6-K9} & Reference \\ & Myr & Excess/Total &
  \% with excess & Excess/Total & \% with excess & \\ \hline 
$\lambda$ Orionis & 5$\pm$1 & 16/44 & 36$^{+48}_{-9}$ & 22/401 &
  5$^{+90}_{-1}$ & \citet{hernandez09,hernandez10} \\
Orion OB1b & 5$\pm$1 & 11/28 & 39$^{+16}_{-12}$ & 3/5$^{X}$ & -- &
\citet{hernandez06} \\ 
Upper Sco & 5$\pm$1 & 10/70 & 14$^{+6}_{-4}$ & 6/23 & 26$^{+16}_{-10}$ &
\citet{carpenter,chen_scocen} \\ 
Orion OB1a & 8.5$\pm$1.5 & 12/25 & 56$^{+19}_{-15}$ & 1/1$^X$ & -- &
\citet{hernandez06} \\ 
$\beta$ Pic & 12$\pm$6 & 5/9$^X$ & -- & 2/12 & 17$^{+29}_{-11}$ &
\citet{rebull} \\
Sco-Cen & 16-17 & 13/25 & 52$^{+19}_{-14}$ & 3/11 &
27$^{+27}_{-15}$ & \citet{chen_scocen} \\ 
\textbf{NGC 1960} & \textbf{16$^{+10}_{-5}$} & \textbf{21/71} &
\textbf{30$^{+24}_{-7}$} & \textbf{14/61} & \textbf{23$^{+69}_{-9}$} &
\textbf{This paper} \\ 
NGC2232 & 25$\pm$4 & 9/23 & 39$^{+26}_{-13}$ & 4/15 & 26$^{+21}_{-13}$ &
\citet{currie_ngc2232} \\ 
IC4665 & 27$\pm$5 & 9/23 & 39$^{+18}_{-13}$ & 12/29 & 41$^{+16}_{-12}$ &
\citet{smith_ic4665} \\ 
NGC2547 & 35$\pm$5 & 15/33 & 45$^{+15}_{-12}$ & 21/41 & 51$^{+14}_{-11}$
& \citet{gorlova_ngc2547} \\ 
IC2391 & 50$\pm$5 & 2/18 & 11$^{+15}_{-7}$ & 6/13 & 46$^{+28}_{-18}$ &
\citet{siegler} \\ 
Blanco 1 & 100$\pm$20 & 4/12 & 33$^{+26}_{-16}$ & 1/25 & 4$^{+9}_{-3}$ &
\citet{stauffer} \\ 
Pleiades & 115$\pm$20 & 7/44 & 16$^{+9}_{-6}$ & 3/38 & 13$^{+9}_{-6}$ &
\citet{gorlova} \\ 
Hyades & 625$\pm$50 & 1/11 & 9$^{+21}_{-8}$ & 1/67 & 1$^{+3}_{-1}$ &
\citet{cieza} \\ 
Field & 4000$^a$ & 0/4$^X$ & -- & 6/65 & 9$^{+6}_{-4}$ &
\citet{bryden} \\ \hline 
\end{tabular}
\begin{flushleft}
\small{Notes: $^X$ Frequency of excess emission in this spectral range
  highly uncertain due to low number of stars, and so this point is
  omitted from Figure \ref{fig:astar_freq} or \ref{fig:solar_freq}. \\
$^a$ Average age for all stars in this paper. }
\end{flushleft}
\end{table*}

\begin{figure*}
\includegraphics[width=16cm]{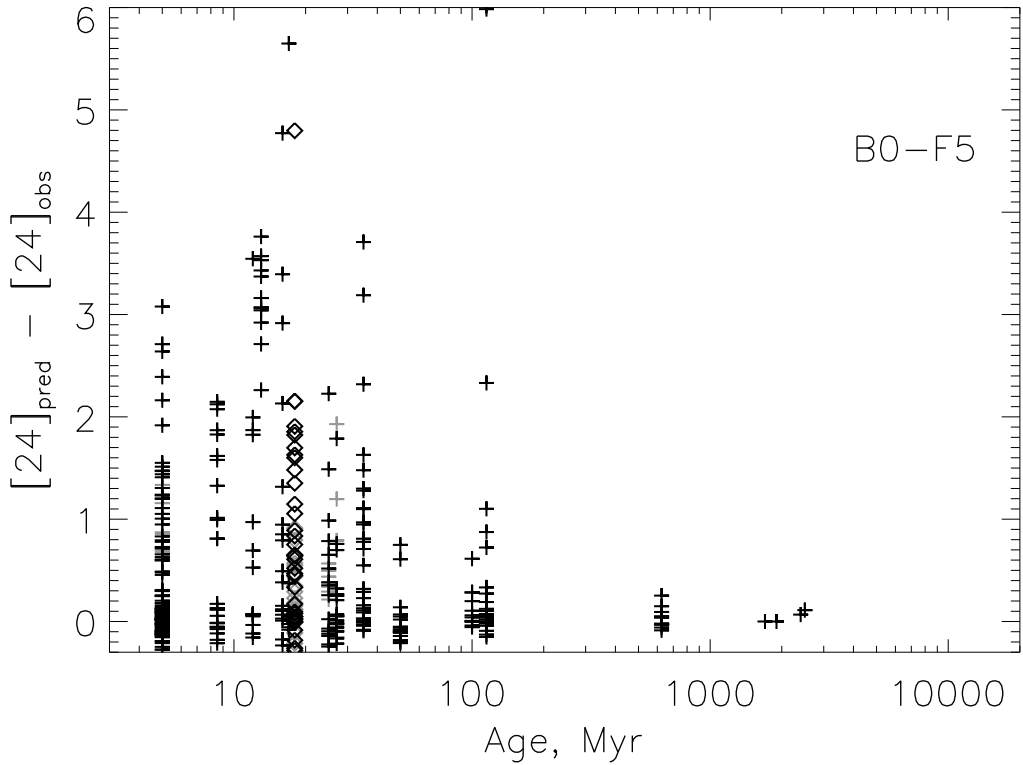}
\caption{\label{fig:atype_xs} The 24$\mu$m excess emission of stars of
  spectral types B0-F5 as a function of age.  Clusters are listed in
  the text.  Data for NGC 1960 are shown by diamonds and offset in age
  to 18 Myr to avoid confusion with the Sco-Cen association.  Grey
  symbols mark 3$\sigma$ upper limits for sources with low signal to
  noise 24$\mu$m data.}
\end{figure*}

In addition to considering the overall frequencies of excess, we also
consider the level of excess emission for each target, given as
$[24]_{\rm{pred}}-[24]_{\rm{obs}}$, where $[24]_{\rm{pred}}$ is
determined from predicted $(K_S - [24])_{\rm{phot}}$ for the target
(see above paragraph).  We show the excesses for stars in the spectral
range B0--F5 in Figure \ref{fig:atype_xs}.  Here we include observed
excess emission from stars in h and $\chi$ Persei
\citep{currie_persei} which were excluded from the statistical
analysis because of uncertain membership lists, and $\beta$ Pictoris
and the field stars which were excluded from Figure
\ref{fig:astar_freq} because of the low number of objects (see Table
\ref{tab:xs_freq}).   Here we again see
that NGC 1960 (shown by diamonds) is comparable to other clusters of a
similar age.  Bright star 14, which has the highest levels of excess
emission in this cluster (and also has excess in the near infrared)
has one of the highest levels of excess emission observed, although
its SED shows that the emission is consistent with debris disc
emission rather than that from a primordial disc (see section 4).
This figure displays more clearly that there is a `rise and fall' in
the upper envelope of 24$\mu$m excess emission, as the peak levels of
24$\mu$m excess reach a maximum at $\sim$10--20Myr before falling off.
A fall-off proportional to time is expected from models of the
evolution of debris discs (e.g. \citealt{dd03,wyattsmith06}).   

\begin{figure}
\includegraphics[width=8cm]{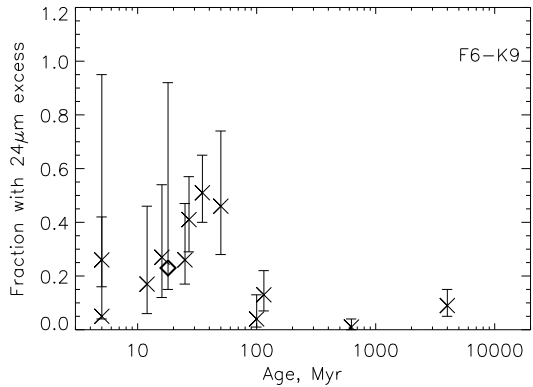}
\caption{\label{fig:solar_freq} The frequency of 24$\mu$m excess
  emission of stars of spectral types F6--K9 as a function of age.
  Clusters are listed in the text and in Table \ref{tab:xs_freq}.
  The datum for NGC 1960 is shown by a diamond and offset in age to 18
  Myr to avoid confusion with the Sco-Cen association. } 
\end{figure}

We show the same plots for lower mass stars of spectral types F6--K9.
In Figure \ref{fig:solar_freq} we can see that the frequency of
24$\mu$m excess falls off roughly in proportion to time after
$\sim$50Myr. NGC 1960 has a fairly typical excess frequency for its
age ($23^{+69}-{9}$\%), although given the large upper limit, we have
really defined a lower limit to the frequency of $>$23\% (excluding
the Poisson statistical errors).  
There is some evidence of the `rise and fall' of debris
discs seen in higher mass stars, although again the statistics for the
younger clusters are highly uncertain.  In this plot it appears that
the peak frequencies of debris discs around F6--K9 type stars are seen
in clusters of ages $\sim 30-50$ Myrs with an abrupt fall-off at
$\sim$100 Myrs (see also Table
\ref{tab:xs_freq}).  This is slower than the expected time taken to
form Pluto-mass proto-planets at $\sim$1 au, and in fact according to
the models of \citet{kenyon05}, a later peak in the 24$\mu$m excess
emission is to be expected for discs at greater distances from the
star (see Figures 7 and 8 of \citealt{kenyon05}).  This suggests that
rather than witnessing excesses arising from terrestrial planet
regions around F6--K9 type stars, we may be seeing the inner edge of
more distant planetesimal belts (due to the apparently smooth increase
in frequencies towards a peak at 30--50 Myrs).  Constraints on longer
wavelength emission for the cluster samples would be required to
explore this possibility.  \citet{siegler} found evidence
that for clusters older than 10Myrs the frequency of 24$\mu$m excess
emission fell-off steadily.  We do not find such a strong relation in
our data (we have considered a larger set of cluster samples),
although it is clear that by 100Myr excess frequencies are 
very low. These differences are likely to arise from different criteria for 
identifying the presence of excess.  This plot demonstrates that the
frequency of debris disc emission is likely to be a function of many
factors aside from the age of the cluster, however with the small
numbers of clusters under consideration, an examination of these
factors is beyond the scope of this paper. 

\begin{figure*}
\includegraphics[width=16cm]{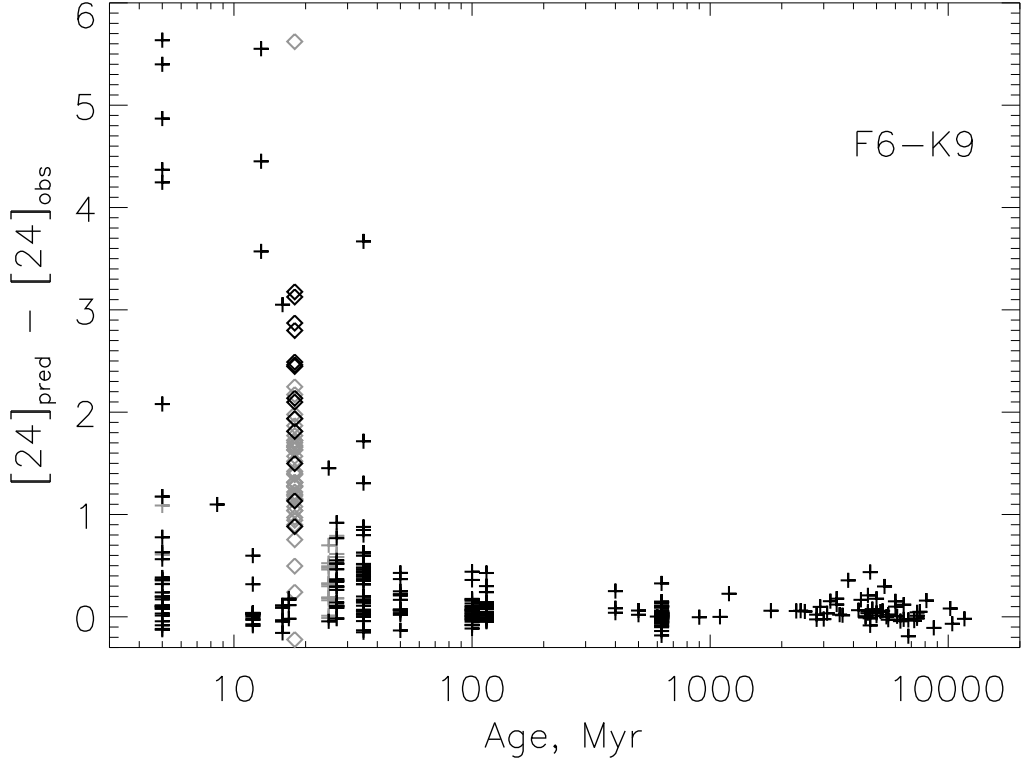}
\caption{\label{fig:solar_xs} The 24$\mu$m excess emission of stars of
  spectral types F6--K9 as a function of age.  Clusters are listed in
  the text.  Data for NGC 1960 are shown by diamonds and offset in age
  to 18Myr to avoid confusion with the Sco-Cen association.  Grey
  symbols mark 3$\sigma$ upper limits for sources with low signal to
  noise 24$\mu$m data.}
\end{figure*}

In Figure \ref{fig:solar_xs} we see that the levels
of excess emission observed around stars in the spectral range F6--K9
in NGC 1960 are typical for stars of similar ages.  We do not see
evidence for a delayed peak (levels of excess emission being higher in
clusters $\sim 10-20$ Myr old) in this spectral range.  This may in part
be due to residual primordial disc populations at 5Myr around lower
mass stars, for example those seen in Upper Sco (sources in this
cluster with $[24]_{\rm{pred}} - [24]_{\rm{obs}} > 4$ were determined to be
primordial in \citealt{carpenter}).  However removal of the primordial
disc targets would mean that the frequency of true debris discs in
these clusters is lower, and therefore the evidence for a delayed peak
would be stronger.  In common with the drop off in
frequency of excess in this spectral range, high levels of excess
emission are not seen at ages $\ge$ 100Myrs in these clusters.  

Interestingly, for NGC 1960 we see no evidence that either the
frequency or the levels of excess emission at 24$\mu$m are dependent on
spectral type.  This is somewhat contrary to the expectation that
24$\mu$m excess should be longer lived around brighter stars, which
arises from the difference in time taken to form planetesimals at
larger radial distances around A-type stars (excess detected at
$\ge$24$\mu$m imply a temperature of 100--150K, translating to an offset of
3--30au around A and early F-type stars and 0.5--3au around solar-type
stars assuming thermal equilibrium).  Within the context of current
planet formation theories \citep{kenyon05, kenyon06}, we might expect
the 24$\mu$m emission around lower mass stars of $\sim$16Myr to be the
result of massive collisions between proto-planetary bodies, and we
might expect it to occur less frequently than the planetesimal
collisions thought to be responsible for the 24$\mu$m emission around
higher mass stars.  In fact, for the
majority of the clusters listed in Table \ref{tab:xs_freq}, there is
no significant difference between the frequency of excess emission
observed around B0--F5-type stars and that observed around F6--K9-type
stars.   However, for the majority of clusters older than $>$5Myr, the
typical levels of excess emission observed around higher mass stars are
higher than that observed around lower mass stars (see Figures
\ref{fig:atype_xs} and \ref{fig:solar_xs}), which does agree with the
model predictions.  This does not appear to be the
case for NGC 1960.  The highest observed excess is indeed observed
around a higher mass star (bright target 14), but in general the
levels of excess emission observed are similar regardless of $V-K_S$
colour (proxy for spectral type).  Our results are biased by the fact
that only relatively high levels of excess will be detected around
lower mass stars, and so there is probably a population of stars in the
range F6--K9 with lower levels of excess than shown in Figure
\ref{fig:solar_xs}.  However, as we do not see a population of higher
mass stars with larger excesses (excluding bright star 14), these
results might suggest that there are a large number of lower mass
sources in the NGC 1960 cluster that are currently undergoing massive
collisions.  The reason why many sources in the cluster should
experience massive collisions at a similar time is unclear, and if
this is the reason for the 24$\mu$m excess observed around lower mass
stars, this presents a challenge to current planet formation theories.  
In Table \ref{tab:highxs} we show the frequencies of stars with large
24$\mu$m excess ($[24]_{\rm{pred}} - [24]_{\rm{obs}} > 2$) for
clusters with populations $>$10 stars in both spectral bins. From
these statistics we cannot see a significant difference between the
frequencies of large excess emission observed in NGC 1960 and other
clusters.  We do find some evidence that the frequency of stars with
large excess emission drops at $\ge$25Myr amongst both high and low
mass samples, which is in agreement with steady-state evolution
models.  We cannot determine if the relationship between the 
frequency of large excesses in the lower mass (F6--K9 type) and higher
mass (B0--F5 type) spectral bins is significantly different for NGC 1960,
partly due to the high errors involved (errors on the
frequencies were determined in the same way as errors on the overall
excess frequencies in Table \ref{tab:xs_freq}). With the current data,
we have no statistically significant evidence that NGC 1960 is
different from other clusters of similar ages.  

\begin{table*}
\caption{\label{tab:highxs} The frequencies of large excess emission
  ($[24]_{\rm{pred}} - [24]_{\rm{obs}} > 2$) among open clusters
  with published 24$\mu$m MIPS photometry.}
\begin{tabular}{lrrcrcl}\hline Name & Age & \multicolumn{2}{c}{B0-F5}
  & \multicolumn{2}{c}{F6-K9} & Reference \\ & Myr & Excess/Total &
  \% with excess & Excess/Total & \% with excess & \\ \hline 
$\lambda$ Orionis & 5$\pm$1 & 0/44 & 0$^{+4}_{-0}$ & 4/401 &
  1$^{+1}_{-1}$ & \citet{hernandez09,hernandez10} \\
Upper Sco & 5$\pm$1 & 2/70 & 3$^{+4}_{-2}$ & 6/23 & 26$^{+16}_{-10}$ &
\citet{carpenter,chen_scocen} \\ 
Sco-Cen & 16-17 & 5/25 & 20$^{+13}_{-9}$ & 1/11 &
9$^{+21}_{-7}$ & \citet{chen_scocen} \\ 
\textbf{NGC 1960} & \textbf{16$^{+10}_{-5}$} & \textbf{3/71} &
\textbf{4$^{+4}_{-2}$} & \textbf{8/61} & \textbf{13$^{+6}_{-4}$} &
\textbf{This paper} \\ 
NGC2232 & 25$\pm$4 & 1/23 & 4$^{+10}_{-3}$ & 0/15 & 0$^{+12}_{-0}$ &
\citet{currie_ngc2232} \\ 
IC4665 & 27$\pm$5 & 0/23 & 0$^{+8}_{-0}$ & 0/29 & 0$^{+6}_{-0}$ &
\citet{smith_ic4665} \\ 
NGC2547 & 35$\pm$5 & 0/33 & 0$^{+6}_{-0}$ & 2/41 & 5$^{+6}_{-3}$
& \citet{gorlova_ngc2547} \\ 
IC2391 & 50$\pm$5 & 0/18 & 0$^{+10}_{-0}$ & 0/13 & 0$^{+14}_{-0}$ &
\citet{siegler} \\ 
Blanco 1 & 100$\pm$20 & 0/12 & 0$^{+15}_{-0}$ & 0/25 & 0$^{+7}_{-0}$ &
\citet{stauffer} \\ 
Pleiades & 115$\pm$20 & 0/44 & 0$^{+4}_{-0}$ & 0/38 & 0$^{+5}_{-0}$ &
\citet{gorlova} \\ 
Hyades & 625$\pm$50 & 0/11 & 0$^{+17}_{-0}$ & 0/67 & 0$^{+3}_{-0}$ &
\citet{cieza} \\ \hline
\end{tabular}
\begin{flushleft}
\small{Notes: Errors on excess frequencies are determined in the same
  manner as for Table \ref{tab:xs_freq} as described in the text.
  Only clusters with 24$\mu$m observations of $>$ 10 stars in each
  spectral bin have been included in this table. }
\end{flushleft}
\end{table*}

In summary the levels and frequency of excess emission seen in NGC
1960 are not significantly different to other open clusters of similar
ages.  The bright target 14 (BD +34 1098) has one of the highest
levels of excess emission observed in early-type stars, but SED
fitting has shown that this is consistent with debris as opposed to
primordial emission.

\subsubsection{Excess emission at 8$\mu$m}

A subset of the open clusters discussed above also have published
8$\mu$m IRAC photometry.  We consider how the frequency of 8$\mu$m
excess evolves with time.  Frequencies are listed by age and spectral
bin in Table \ref{tab:8xs_freq}.   Errors on the frequency are derived
from upper limits and the Poisson statistics of \citet{gehrels} as
described for the 24$\mu$m data in the preceeding section.

\begin{table*}
\caption{\label{tab:8xs_freq} The excess frequencies of open clusters
  with published 8$\mu$m IRAC photometry.}
\begin{tabular}{lrrcrcl}\hline Name & Age & \multicolumn{2}{c}{B0-F5}
  & \multicolumn{2}{c}{F6-K9} & Reference \\ & Myr & Excess/Total &
  \% with excess & Excess/Total & \% with excess & \\ \hline 
$\lambda$ Orionis & 5$\pm$1 & 2/44 & 5$^{+6}_{-3}$ & 15/401 & 4$^{+1}_{-1}$ &
  \citet{hernandez09, hernandez10} \\
Orion OB1b & 5$\pm$1 & 1/29 & 3$^{+34}_{-3}$ & 0/5$^X$ & -- &
\citet{hernandez06} \\
Upper Sco & 5$\pm$1 & 5/68 & 7$^{+5}_{-3}$ & 4/45 & 9$^{+8}_{-5}$ &
\citet{carpenter06} \\ 
Orion OB1a & 8.5$\pm$1.5 & 2/25 & 8$^{+12}_{-5}$ & 0/1$^X$ & -- &
\citet{hernandez06} \\
\textbf{NGC 1960} & \textbf{16$^{+10}_{-5}$} & \textbf{3/71} &
\textbf{4$^{+4}_{-2}$} & \textbf{0/61} & \textbf{$<$7} & \textbf{This
  paper} \\
NGC 2232 & 25$\pm$4 & 1/23 & 4$^{+10}_{-4}$ & 1/15 & 7$^{+15}_{-6}$ &
\citet{currie_ngc2232} \\
IC4665 & 27$\pm$5 & 1/23 & 4$^{+10}_{-4}$ & 0/29 & $<$10 & \citet{smith_ic4665}
\\
NGC 2547 & 35$\pm$5 & 2/33 & 6$^{+8}_{-4}$ & 2/41 & 5$^{+6}_{-3}$ &
\citet{gorlova_ngc2547} \\ 
\hline 
\end{tabular}
\begin{flushleft}
\small{Notes: $^X$ Frequency of excess emission in this spectral range
  highly uncertain due to low number of stars. \\
$^a$ No significant detections at 8$\mu$m in this spectral range. }
\end{flushleft}
\end{table*}

These figures confirm that 8$\mu$m excess emission is rare amongst
stars of ages $\ge$5Myr.   We do not see any evidence for a dependence
of 8$\mu$m excess frequency on the age of the cluster.  As we only have
upper limits for many of the clusters, particularly for the lower
mass spectral bin, any dependence of 8$\mu$m frequency on age or
spectral type is
difficult to determine.  However we can state that the 8$\mu$m excess
frequency is always lower than the 24$\mu$m frequency.  This
demonstrates that hot--warm dust is much less common than warm--cool
debris disc emission.  This in turn suggests that discs are likely to
lie at a few to tens of au from the star, and not within 1 or a few
au.  This is consistent with the picture of debris discs arising from
statistical studies and studies of individual debris disc hosts (see
\citealt{wyattreview} and references therein).  

\section{Conclusions}

In this paper we have presented a study of the cluster NGC 1960 (M36)
using archival Spitzer MIPS and IRAC data.  These data have been used
to search for debris discs in the cluster.  Our conclusions are as
follows: 

\begin{enumerate}
\item We have identified 38 targets as having significant 24$\mu$m
  excess.  One target, bright star 14 (BD +34 1098), has very high
  levels of excess ($F_{24}/F_{\rm{phot}} \sim$85).  Three targets
    (including BD +34 1098) have significant excess emission at
    near-infrared wavelengths.  The SEDs of these targets have been
    examined and the emission has been determined to be consistent
    with second generation debris emission, as opposed to remnant
    primordial discs.
\item The frequency and overall levels of excess emission observed in
  NGC 1960 are consistent with other clusters of similar ages ($>$30\%
  of B0--F5 type stars, and $>$23\% of F6-K9 stars have significant
  excess).  The data from this cluster are consistent with a `rise and
  fall' in debris disc emission amongst higher mass stars, and with a
  possible delayed peak in emission amongst lower mass stars. 
\item We find no evidence for a dependence of excess emission on
  stellar multiplicity.  We are however restricted in our examination
  of this relationship by a lack of completeness in the 24$\mu$m flux
  measurements for intermediate to lower mass stars.  
\end{enumerate}

\section*{Acknowledgments}

This work is based on archival data obtained with the Spitzer
Space Telescope, which is operated by the Jet Propulsion Laboratory,
California Institute of Technology under a contract with NASA.  This
paper has made use of data obtained at the William Herschel Telescope,
which is operated on the island of La Palma by the Isaac Newton Group
in the Spanish Observatorio del Roque de los Muchachos of the
Instituto de Astrofísica de Canarias. R.S. would like to acknowledge
the support of the UK Science and Technology Facilities Council
(STFC).

\bibliographystyle{mn2e}  
\bibliography{/home/rachel/Work/PAPERs/thesis} 

\label{lastpage}

\end{document}